\documentclass[11pt,a4paper]{article}
\pdfoutput=1
\usepackage{jheppub}
\usepackage{amsfonts,amsbsy,amsmath,amssymb}
\usepackage{graphicx}
\usepackage{booktabs}
\usepackage{caption}
\usepackage{subcaption}
\usepackage{multirow}
\title{Symanzik improvement of the gradient flow in lattice gauge theories}

\author[a]{Alberto Ramos}
\author[b]{and Stefan Sint}

\affiliation[a]{PH-TH, CERN, CH-1211 Geneva 23, Switzerland}
\affiliation[b]{School of Mathematics, Trinity College Dublin, Dublin
  2, Ireland.}
\emailAdd{alberto.ramos@cern.ch}
\emailAdd{sint@maths.tcd.ie}

\abstract{
We apply the Symanzik improvement programme to the  4+1-dimensional local
re-formulation of the gradient flow in pure $SU(N)$ lattice gauge theories.
We show that the classical nature of the flow equation allows to eliminate all
cutoff effects at $\mathcal O(a^2)$ which originate either from the
discretized 
gradient 
flow equation or from the gradient flow observable. 
All the remaining $\mathcal O(a^2)$ effects can be understood in terms of
local counterterms at the zero flow time boundary.
We classify these counterterms and provide a complete set as required
for on-shell improvement. Compared to the 4-dimensional pure gauge theory
only a single additional counterterm is required, 
which corresponds to a modified initial condition for the flow equation.
A consistency test in perturbation theory is passed and allows to
determine all 
counterterm coefficients to lowest non-trivial order in the coupling.
}

\preprint{
  \begin{flushright}
    CERN-PH-TH-2015-199\\
    TCDMATH 15--06
  \end{flushright}
}
\newcommand{\unit}{1\kern-.25em {\rm l}}
\newcommand{\tr}{{\rm tr}}
\newcommand{\rmO}{{\rm O}}
\begin{document}
\maketitle

\section{Introduction}

In recent years the Yang-Mills gradient flow has been established as
a very promising new tool to study non-perturbative aspects of
strongly coupled 
gauge
theories~\cite{Narayanan:2006rf,Luscher:2010iy,Luscher:2011bx,Luscher:2013cpa}.
The gradient flow defines a deterministic mapping from the
original gauge field $A_\mu(x)$ to a smoothed gauge field configuration,
$B_\mu(t,x)$, at flow time $t$, which is obtained as the solution
of the gradient flow equation (see appendix~\ref{ap:conv} for a
summary of our conventions),
\begin{equation}
 \partial_t B_\mu(t,x) = \sum_{\nu} D_\nu G_{\nu\mu}(t,x),\qquad
 B_\mu(0,x) = A_\mu(x), 
 \label{eq:YMflow}
\end{equation}
where $D_\mu = \partial_\mu + [B_\mu,\cdot]$ denotes the gauge
covariant derivative and 
\begin{equation}
  G_{\mu\nu}= \partial_\mu B_\nu - \partial_\nu B_\mu + [B_\mu,B_\nu],
\end{equation}
is the associated field strength tensor. 
The name relates to the fact that the right hand side of (\ref{eq:YMflow})
is equal to minus the gradient of the Yang-Mills gauge action.
Hence, with increasing flow time $t$, the solution, $B_\mu(t,x)$,
is driven towards a minimum of the action and thus approaches
a smooth classical field configuration.

There is quite some freedom when translating the gradient flow equation
to a Euclidean space-time lattice.
A simple possibility is to choose Wilson's plaquette action, $S_{\text{W}}$,
and to define the lattice gauge field at finite flow time, $V_\mu(t,x)$, as
the solution of the Wilson flow equation,
\begin{equation}
  a^{2} \left[\partial_t V_\mu(t,x)\right]V_\mu(t,x)^\dagger
  = -g_0^2\partial_{x,\mu} S_{\text{W}}[V],
  \label{eq:wflow}
\end{equation}
where $\partial_{x,\mu}$ denotes the Lie-algebra
valued derivative
with respect to $V_\mu(t,x)$. It should be noted that similar smoothing
operations have long been successfully applied in lattice QCD.
For example, the stout link smearing technique of
ref.~\cite{Morningstar:2003gk} 
can be understood as discretized flow time version of Eq.~(\ref{eq:wflow}),
The essential new element is a theoretical understanding of the
renormalization 
properties of the Yang-Mills gradient flow. In particular,
in~\cite{Luscher:2011bx,Luscher:2013cpa}
it was proved to all orders of perturbation theory that QCD at finite
flow time $t$ is 
renormalized once it is renormalized at flow time $t=0$ through the usual 
renormalizations of the gauge coupling and the quark mass parameters.
Furthermore, gauge invariant fields at positive flow time are automatically
renormalized and do not mix with other fields of the same or lower
dimensions.  These properties allow to define a new class of renormalized
gauge invariant observables which can be used to probe the theory
in various ways. It also opens new ways to define renormalized
composite operators at zero flow time; the study of Ward identities at
positive flow
times~\cite{Luscher:2013cpa,DelDebbio:2013zaa,Shindler:2013bia} and  
the applications of the so called ``small flow time expansion'' have
received much attention  recently in this
context~\cite{Suzuki:2013gza,Asakawa:2013laa,Makino:2014taa}. 

Many current lattice QCD applications of the gradient flow only
involve the simplest possible gauge invariant field, the action density,
\begin{equation}
  \label{eq:actden}
E(t,x)= -\frac12\sum_{\mu,\nu} \tr\{ G_{\mu\nu}(t,x)G_{\mu\nu}(t,x)\}.
\end{equation}
As initially proposed in~\cite{Luscher:2010iy}, the expectation value
$\langle E(t,x)\rangle$ can be used for a non-perturbative definition
of either a reference  scale or a coupling constant. This has proven
very attractive: in 
large volume simulations  it leads to the most precise determination
of a reference scale 
(for a recent review cf.~\cite{Sommer:2014mea}). On the other hand, when
considered in a finite space-time volume the scale evolution of the
corresponding
coupling~\cite{Fodor:2012td,Fritzsch:2013je,Luscher:2014kea,Ramos:2014kla,GarciaPerez2015}  
can be traced with high statistical precision
(see~\cite{Ramos:2015dla} for a recent review).

Notwithstanding these nice properties a major practical problem are
the relatively large  cutoff effects which have been observed in several
applications (cf.~\cite{Ramos:2015dla} and references therein).
On general grounds, the leading effects are expected to be of order $a^2$.
Their size depends on the detailed choices made when translating the flow
equation (\ref{eq:YMflow}) to the lattice, but also on the
discretization of the 
observable and on the lattice action. 
Alternative flow equations have been tried, e.g.~in
ref.~\cite{Borsanyi:2012zs} 
where the Wilson action was replaced by the tree-level improved
L\"uscher-Weisz action, 
$S_{\rm LW}$~\cite{Luscher:1984xn,Luscher:1985zq}. For some attempts
to reduce cutoff effects in the particular observable
$\langle E(t,x)\rangle$ cf.~refs.~\cite{Cheng:2014jba,Fodor:2014cpa}.
Here we would like to proceed more systematically by applying the
Symanzik procedure~\cite{Symanzik:1983dc,Luscher:1984xn} to the 
4+1-dimensional local formulation of the
theory~\cite{ZinnJustin:1987ux,Luscher:2011bx}. 
This will lead us to a particular choice for the lattice flow equation,
referred to as the ``Zeuthen flow" and defined by 
\begin{equation}
  \label{eq:impflowlat}
  a^2\left(\partial_t V_\mu(t,x)\right) V_\mu(t,x)^\dagger  = -g_0^2 \left(1 +
    \frac{a^2}{12}\nabla_\mu^\ast\nabla_\mu^{}
  \right) \partial_{x,\mu} S_{\rm LW}[V] \,, 
  \qquad V_\mu(0,x) = U_\mu(x)  \,.
\end{equation}
Here $\nabla_\mu^{}$ and $\nabla_\mu^\ast$ are the lattice forward and
backward covariant derivatives, respectively. We will show that the
integration of the Zeuthen flow equation does not generate 
any cutoff effects at O($a^2$). If combined with classical O($a^2$)
improvement of the observable all O($a^2$) effects are eliminated
apart from those corresponding to local counterterms in the action at
zero flow time. We will give a complete list of such counterterms and
test our framework to lowest non-trivial order in perturbation theory.

The paper is organized as follows: In Section~2 we recall the definition
of the 4+1-dimensional local theory, with flow time as the added dimension.
In Section~3 we discuss the general Symanzik procedure and the
simplifications 
due to the special properties of this theory.
We present the classical $a$-expansion of both the flow action and the
gradient flow 
observable $E(t,x)$, as part of the simplified Symanzik procedure, and carry
out the standard Symanzik analysis for the O($a^2$) counterterms at
the $t=0$ boundary. 
Section~4 presents a number of perturbative tests of the O($a^2$)
improved theory, 
and Section~5 our conclusions. We have included three appendices regarding
our notations and conventions (Appendix A),
some technical details pertaining to the classical $a$-expansion
(Appendix~\ref{ap:odd}), 
and some explicit expressions used in Section~4 (Appendix C), respectively.


\section{Lattice gauge theory in 4+1 dimensions}

The gradient flow equation can be viewed as a way to define
a particular class of observables,
i.e.~fields which are functionals of the fundamental gauge field $U_\mu(x)$.
The flow time thus appears as an additional parameter which measures
the range in space-time over which the fundamental
gauge field enters into an observable defined in terms of the
flowed gauge field $V_\mu(t,x)$.
The flow time $t$ has dimension length squared and the ``smearing radius"
$r_t=\sqrt{8t}$ is usually taken as the corresponding
length scale\footnote{The radius $r_t=\sqrt{8t}$
amounts to 2 standard deviations in the Gaussian smearing function
which appears in the relation between $B_\mu(t,x)$ and $A_\mu(x)$
to leading order in the coupling.}. 
Thus, gradient flow observables are non-local objects from the perspective
of the 4-dimensional gauge theory and their properties under
renormalization are difficult to assess. Moreover, the non-locality
prevents a 
straightforward application of the Symanzik expansion, which
is our main theoretical tool for understanding
the cutoff dependence of the theory. For this purpose, it is therefore
highly beneficial to  follow~\cite{Luscher:2013cpa} and view the
theory from a 
4+1-dimensional perspective, with flow time as the added dimension. In
this re-formulation locality is restored in the 
4+1-dimensional sense, and dimensional counting can be applied to
classify counterterms to the action and observables.

We start with the formulation of the lattice set-up, including
the introduction of a flow-time lattice. The latter should be regarded
as an intermediate regularization which helps to resolve certain technical
issues~\cite{Luscher:2013cpa}. While none of this is original it serves for 
later reference and to fix our notation.

\subsection{The 4-dimensional lattice action}

On-shell O($a^2$) improvement of the 4-dimensional gauge theory can be
achieved by  
introducing, besides the 4-link plaquette action, further 6-link
Wilson loops with 
appropriately chosen coefficients~\cite{Luscher:1984xn}.
We will consider a general class of lattice gauge actions 
parameterized by the coefficients $c_i (i=0,1,2,3)$, defined by,
\begin{equation}
  \label{eq:latac}
  S_{\text{g}}[U,\{c_i\}] = \frac{1}{g_0^2} \sum_{i=0}^3 c_i
  \sum_{\mathcal W\in \mathcal 
    S_i} {\rm Tr}(1 - U(\mathcal C))\,,  
\end{equation}
where the second sum extends over all oriented Wilson loops of type $\mathcal
S_i$. As illustrated in Fig.~\ref{fig:loops}, these Wilson loops are
the usual plaquettes, $\mathcal S_0$, the $2\times 1$ planar loops or
``rectangles", $\mathcal S_1$, the 
bent rectangles or ``chairs", $\mathcal S_2$, and finally the
``parallelograms", $\mathcal S_3$. 
\newcommand{\basispl}{
   \put(-.5,-.5){\line(1,0){1}}
   \put(.5,-.5){\line(0,1){1}}
   \put(.5,.5){\line(-1,0){1}}
   \put(-.5,.5){\line(0,-1){1}}
                         }
\newcommand{\basisar}{
   \put(0,-.5){\vector(1,0){0}}
   \put(.5,0){\vector(0,1){0}}
   \put(0,.5){\vector(-1,0){0}}
   \put(-.5,0){\vector(0,-1){0}}
	              }
\newcommand{\revar}{
   \put(0,-.5){\vector(-1,0){0}}
   \put(.5,0){\vector(0,-1){0}}
   \put(0,.5){\vector(1,0){0}}
   \put(-.5,0){\vector(0,1){0}}
	              }
\newcommand{\plaq}{\setlength{\unitlength}{.5cm}\raisebox{-.2cm}{
   \begin{picture}(1.2,1.2)(-.6,-.6)
   \basispl\basisar
   \put(-.5,-.5){\circle*{.2}}
   \put(-.55,-.55){\makebox(0,0)[tr]{\footnotesize $x$}}
   \put(-.55,0){\makebox(0,0)[r]{\footnotesize $\nu$}}
   \put(0,-.55){\makebox(0,0)[t]{\footnotesize $\mu$}}
   \end{picture}}}
\newcommand{\twoplaq}{\setlength{\unitlength}{1cm}\raisebox{-.5cm}{
   \begin{picture}(1.2,1.2)(-.6,-.6)
   \basispl
   \put(-.5,-.5){\circle*{.1}}
   \put(-.5,.5){\circle*{.1}}
   \put(.5,-.5){\circle*{.1}}
   \put(.5,.5){\circle*{.1}}
   \put(0,-.5){\circle*{.1}}
   \put(0,.5){\circle*{.1}}
   \put(.5,0){\circle*{.1}}
   \put(-.5,0){\circle*{.1}}
   \put(-.25,-.5){\vector(1,0){0}}
   \put(.25,-.5){\vector(1,0){0}}
   \put(.5,-.25){\vector(0,1){0}}
   \put(.5,.25){\vector(0,1){0}}
   \put(-.25,.5){\vector(-1,0){0}}
   \put(.25,.5){\vector(-1,0){0}}
   \put(-.5,-.25){\vector(0,-1){0}}
   \put(-.5,.25){\vector(0,-1){0}}
   \put(-.55,-.55){\makebox(0,0)[tr]{\footnotesize $x$}}
   \put(-.55,0){\makebox(0,0)[r]{\footnotesize $\nu$}}
   \put(0,-.55){\makebox(0,0)[t]{\footnotesize $\mu$}}
   \end{picture}}}
\newcommand{\stapup}{\setlength{\unitlength}{.5cm}\raisebox{-.2cm}{
   \begin{picture}(1.2,1.2)(-.6,-.6)
   \put(.5,-.5){\line(0,1){1}}
   \put(.5,.5){\line(-1,0){1}}
   \put(-.5,.5){\line(0,-1){1}}
   \put(.5,0){\vector(0,-1){0}}
   \put(0,.5){\vector(1,0){0}}
   \put(-.5,0){\vector(0,1){0}}
   \put(-.5,-.5){\circle*{.2}}
   \put(-.55,-.55){\makebox(0,0)[tr]{\footnotesize $x$}}
   \put(-.55,0){\makebox(0,0)[r]{\footnotesize $\nu$}}
   \put(0,.55){\makebox(0,0)[b]{\footnotesize $\mu$}}
   \end{picture}}}
\newcommand{\stapdw}{\setlength{\unitlength}{.5cm}\raisebox{-.2cm}{
   \begin{picture}(1.2,1.2)(-.6,-.6)
   \put(.5,-.5){\line(0,1){1}}
   \put(.5,-.5){\line(-1,0){1}}
   \put(-.5,.5){\line(0,-1){1}}
   \put(.5,0){\vector(0,1){0}}
   \put(0,-.5){\vector(1,0){0}}
   \put(-.5,0){\vector(0,-1){0}}
   \put(-.5,.5){\circle*{.2}}
   \put(-.55,.75){\makebox(0,0)[tr]{\footnotesize $x$}}
   \put(-.55,0){\makebox(0,0)[r]{\footnotesize $\nu$}}
   \put(0,-.55){\makebox(0,0)[t]{\footnotesize $\mu$}}
   \end{picture}}}
\newcommand{\clover}{\setlength{\unitlength}{.5cm}\raisebox{-.5cm}{
   \begin{picture}(2.4,2.4)(-1.2,-1.2)
   \multiput(-1.2,-1.2)(1.2,1.2){2}{\begin{picture}(1.2,1.2)(-.6,-.6)
   \basispl\basisar\end{picture}}
   \multiput(-1.2,0)(1.2,-1.2){2}{\begin{picture}(1.2,1.2)(-.6,-.6)
   \basispl\revar\end{picture}}
   \put(-.1,-.1){\circle*{.2}}
   \put(-.1,.1){\circle*{.2}}
   \put(.1,-.1){\circle*{.2}}
   \put(.1,.1){\circle*{.2}}
   \end{picture}}}
\newcommand{\twooneplaq}{\setlength{\unitlength}{.5cm}
   \raisebox{-.2cm}{
   \begin{picture}(2.2,1.2)(-1.1,-.6)
   \put(-1,-.5){\line(1,0){2}}
   \put(-1,.5){\line(1,0){2}}
   \put(-1,-.5){\line(0,1){1}}
   \put(1,-.5){\line(0,1){1}}
   \multiput(-1,-.5)(1,0){3}{\circle*{.2}}
   \multiput(-1,.5)(1,0){3}{\circle*{.2}}
   \end{picture}}}
\newcommand{\plaqa}{\setlength{\unitlength}{.5cm}\raisebox{-.2cm}{
   \begin{picture}(1.2,1.2)(-.6,-.6)
   \basispl
   \put(-.5,-.5){\circle*{.2}}
   \put(-.5,.5){\circle*{.2}}
   \put(.5,-.5){\circle*{.2}}
   \put(.5,.5){\circle*{.2}}
   \end{picture}}}
\newcommand{\hookplaq}{\setlength{\unitlength}{.5cm}
   \raisebox{-.3268cm}{
   \begin{picture}(1.7071,1.7071)(-.7071,-.7071)
   \put(0,0){\line(0,1){1}}
   \put(0,1){\line(1,0){1}}
   \put(1,1){\line(0,-1){1}}
   \put(-.7071,-.7071){\line(1,0){1}}
   \put(0,0){\line(-1,-1){.7071}}
   \put(1,0){\line(-1,-1){.7071}}
   \multiput(0,0)(1,0){2}{\circle*{.2}}
   \multiput(0,1)(1,0){2}{\circle*{.2}}
   \multiput(-.7071,-.7071)(1,0){2}{\circle*{.2}}
   \multiput(0,0)(.25,0){4}{\circle*{.03}}
   \end{picture}}}
\newcommand{\cornplaq}{\setlength{\unitlength}{.5cm}
   \raisebox{-.3268cm}{
   \begin{picture}(1.7071,1.7071)(-.7071,-.7071)
   \put(-.7071,-.7071){\line(0,1){1}}
   \put(0,1){\line(1,0){1}}
   \put(1,1){\line(0,-1){1}}
   \put(-.7071,-.7071){\line(1,0){1}}
   \put(0,1){\line(-1,-1){.7071}}
   \put(1,0){\line(-1,-1){.7071}}
   \put(-.7071,-.7071){\circle*{.1}}
   \put(-.7071,.2929){\circle*{.2}}
   \multiput(0,0)(1,0){2}{\circle*{.2}}
   \multiput(0,1)(1,0){2}{\circle*{.2}}
   \multiput(-.7071,-.7071)(1,0){2}{\circle*{.2}}
   \multiput(0,0)(.25,0){4}{\circle*{.03}}
   \multiput(0,0)(0,.25){4}{\circle*{.03}}
   \multiput(0,0)(-.1768,-.1768){4}{\circle*{.03}}
   \end{picture}}}


\begin{figure}
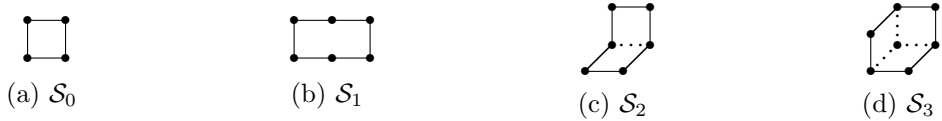

  \centering
  \begin{subfigure}[t]{0.24\textwidth}
    \centering
    \plaqa
    \caption{$\mathcal S_0$}
  \end{subfigure}
  \begin{subfigure}[t]{0.24\textwidth}
    \centering
    \twooneplaq
    \caption{$\mathcal S_1$}
  \end{subfigure}
  \begin{subfigure}[t]{0.24\textwidth}
    \centering
    \hookplaq
    \caption{$\mathcal S_2$}
  \end{subfigure}
  \begin{subfigure}[t]{0.24\textwidth}
    \centering
    \cornplaq
    \caption{$\mathcal S_3$}
  \end{subfigure}

  \caption{Wilson loops of type $\mathcal S_0, \mathcal S_1, \mathcal S_2$ and $\mathcal S_3$.}
  \label{fig:loops}
\end{figure}
It is conventional to normalize the action by requiring
\begin{equation}
 c_0+ 8c_1+16 c_2 +8 c_3 =1\,,
\end{equation}
such that the standard continuum Yang-Mills action is obtained in the
classical continuum limit, with any choice of the
3 free parameters\footnote{Note, however, that the choice of the
  coefficients is not completely free,  
with some constraints arising from
positivity~\cite{Luscher:1984xn}. Our conventions differ from this
reference by  
the exchange~$c_2 \leftrightarrow c_3$.}.
Popular choices are the Wilson plaquette (W) action ($c_0=1$, $c_{1,2,3}=0$)
and the tree-level improved L\"uscher-Weisz (LW) action ($c_0=5/3$, $c_1=-1/12$, $c_{2,3}=0$).

\subsection{4+1-dimensional set-up}

Given the 4-dimensional action, the flow equation is now incorporated 
in the action as a constraint, by introducing the Lagrange multiplier field
$L_\mu(t,x)$, which is hermitian and such that $\imath L_\mu(t,x)$ is
Lie-algebra valued. 
The 4+1-dimensional action of this theory then takes the form
\begin{equation}
  \label{eq:act4plus1}
  S[V,L] = S_{\text{g}}[U,\{c_i\}] - 2a^4\int_0^\infty dt\,
  \sum_{x,\mu} \tr\{L_\mu(t,x) F_\mu(t,x)\}, 
\end{equation}
where the boundary condition,
\begin{equation}
  V_\mu(0,x) = U_\mu(x),
\end{equation}
is assumed and
\begin{equation}
 \label{eq:Fmudef}
  F_\mu(t,x) =  a^{-1} \left(\partial_t V_\mu(t,x)\right)V_\mu(t,x)^\dagger
  -  a^{-3} g_0^2 \partial_{x,\mu} S_{\text{g}}[V],
\end{equation}
is a shorthand notation which allows to write the lattice gradient
flow equation in the form $F_\mu(t,x)=0$.  The action $S_{\text{g}}[V]$
is some 4-dimensional lattice gauge action for the flowed field $V_\mu(t,x)$,
the Wilson action being the simplest choice [cf. eq.~(\ref{eq:wflow})]. In any case
it is unrelated to the gauge action $S_{\text{g}}[U]$ in (\ref{eq:act4plus1}).
How to best define $F_\mu(t,x)$ is at the core of this work and
will be discussed in the next section.

Given the action for the 4+1-dimensional
half space $t\ge 0$,  expectation values of composite fields $O[V,L]$
are defined as usual, 
\begin{equation}
  \label{eq:Z4p1d}
  \langle O\rangle = {\cal Z}^{-1}\int D[V] D[L] O[V,L] \exp\left(-
    S[V,L]\right),\qquad \langle 1 \rangle =1\,.
\end{equation}
A few remarks are in order: first, the integration over the gauge
field $V_\mu(t,x)$ includes the integration over its boundary values at $t=0$
i.e.~the standard 4-dimensional gauge field $U_\mu(x)$.
Hence, for observables which only depend on $U_\mu$, the functional integrals
over $V_\mu\vert_{t>0}$ and $L_\mu$ cancel between numerator and denominator, reproducing the
standard expectation value of the 4-dimensional theory.
To see this more explicitly it is convenient to pass to a flow time lattice with 
spacing $\varepsilon$ and lattice points
$t=n\varepsilon$~\cite{Luscher:2013cpa}, 
\begin{equation}
\label{eq:Sbulkeps}
  \int_0^\infty dt\, a^4\sum_{x,\mu} \tr\{L_\mu(t,x) F_\mu(t,x)\} \longrightarrow
  \varepsilon \sum_{n\ge 0} a^4\sum_{x,\mu} \tr\{L_\mu(t,x) F_\mu(t,x;\varepsilon)\},
\end{equation}
where we have assumed the discretization, 
\begin{equation}
  \label{eq:Fmueps}
  a\varepsilon F_\mu(t,x;\varepsilon) = V_\mu(t+\varepsilon,x) V_\mu(t,x)^\dagger
  - \exp\left(-g_0^2\frac{\varepsilon}{a^2}\partial_{x,\mu} S_{\text{g}}[V]\right),
\end{equation}
with the correct $\varepsilon \rightarrow 0$ limit. Inserting this representation 
of the action into the functional integral, the integration over the
fields $L_\mu(t,x)$ produces a string of $\delta$-functions\footnote{For a more careful discussion of the limits
involved cf.~\cite{Luscher:2013cpa}}
\begin{equation}
   \prod_{x,\mu}\prod_{n=0}^\infty \delta\left[F_\mu(n\varepsilon,x;\varepsilon)\right].
\end{equation}
These can be eliminated one by one, by integrating over $V_\mu(n\varepsilon,x)$
for strictly positive $n$, leaving the unconstrained $n=0$ integration
over the fundamental gauge field intact, as expected.

\subsection{Gauge symmetry}
\label{subsect:gaugesym}

By construction, the 4+1-dimensional action is gauge invariant under $t$-independent
gauge transformations,
\begin{equation}
  V_\mu(t,x) \rightarrow \Lambda(x) V_\mu(t,x) \Lambda(x+a\hat\mu)^\dagger,
\end{equation}
where $\Lambda(x)$ is an SU($N$)-valued gauge function. This leads to
the transformation,
\begin{equation}
  F_\mu(t,x) \rightarrow \Lambda(x)  F_\mu(t,x) \Lambda(x)^\dagger,
\end{equation}
so that gauge invariance of the action is guaranteed provided that
\begin{equation}
   L_\mu(t,x) \rightarrow \Lambda(x)  L_\mu(t,x) \Lambda(x)^\dagger,
\end{equation}
i.e.~$L_\mu(t,x)$ must be in the adjoint representation of the
gauge group.  The integration measure is invariant under such a change of variables,
so that  the gauge symmetry of the 4-dimensional boundary theory is inherited by
the bulk theory.

It is occasionally useful to generalize the gauge symmetry to the
flow-time coordinate $t$, i.e.~admit flow-time dependent
gauge functions $\Lambda(t,x)$.
In the continuum theory this amounts to replacing $t=x_4$,
$\partial_t \rightarrow D_4=\partial_4 + [B_4,\cdot]$ and $\partial_t B_\mu(t,x) \rightarrow
G_{4\mu}(x_4,x)$~\cite{Narayanan:2006rf}. In the presence of the lattice cutoff
(but continuous $t=x_4$) we define the covariant $x_4$-derivative by
\begin{equation}
  \nabla_4 V_\mu(\tilde{x}) =
  \partial_4 V_\mu(\tilde{x}) + B_4(\tilde{x}) V_\mu(\tilde{x}) -
  V_\mu(\tilde{x})B_4(\tilde{x}+a\hat\mu),
\end{equation}
where $\tilde{x} = (x_4,x)$. This, together with the transformation
under an $x_4$-dependent gauge transformation 
\begin{equation}
  B_4(\tilde{x}) \rightarrow \Lambda(\tilde{x}) B_4(\tilde{x})  \Lambda(\tilde{x})^\dagger
+ \Lambda(\tilde{x})\partial_4 \Lambda(\tilde{x})^\dagger,
\end{equation}
leads to the left hand side of the covariant flow equation transforming as
\begin{equation}
  \left[\nabla_4 V_\mu(\tilde{x})\right] V_\mu(\tilde{x})^\dagger \rightarrow
  \Lambda(\tilde{x})\left[\nabla_4 V_\mu(\tilde{x})\right]
  V_\mu(\tilde{x})^\dagger \Lambda(\tilde{x})^\dagger\,.
\end{equation}
Rendering the $t$-derivative covariant in the definition of $F_\mu$ (\ref{eq:Fmudef}) one
then obtains,
\begin{equation}
  F_\mu(\tilde{x}) \rightarrow \Lambda(\tilde{x})  F_\mu(\tilde{x}) \Lambda(\tilde{x})^\dagger,
\end{equation}
so that 4+1-dimensional gauge invariance is established, provided
that $L_\mu(\tilde{x})$ transforms just like $F_\mu(\tilde{x})$.
Discretizing the flow-time coordinate is also straightforward,
one just needs to elevate the fourth component
of the gauge field $B_4$ to a link field $V_4$, with corresponding changes
in the covariant derivative and gauge tranformation behaviour.

Finally we note that the Yang-Mills flow equation in the continuum can
be written as 
\begin{equation}
  G_{4\mu} = \sum_{\nu=0}^3 D_\nu G_{\nu\mu},
\end{equation}
which shows that the 4+1-dimensional theory, while exactly gauge
invariant, does not enjoy any generalized Lorentz-symmetry.
This is of course already clear from the
dimensions, in particular, $\partial_t$ and thus $B_4$ must have mass
dimension $2$, in contrast to the usual derivatives and gauge fields in 4 dimensions.


\section{Symanzik improvement to $\mathcal O(a^2)$}

\subsection{Generalities}

The re-formulation of gradient flow observables in terms of 
a local 4+1-dimensional lattice gauge theory creates the standard situation 
to which Symanzik's effective theory~\cite{Symanzik:1983dc} can be
applied in the usual way. 
We start with Symanzik's effective action which is given as an
expansion in powers 
of $a^2$, 
\begin{equation}
  \label{eq:symact}
   S_\text{eff}[B,L] = S_0^{\text{cont}}[B,L] + a^2 S_{2,{\rm fl}}[B,L] + a^2
   S_{2,b}[B,L] + O(a^4). 
\end{equation}
One might worry about odd powers of $a$ arising in a 4+1-dimensional theory.
However, as we will show in detail in Appendix~\ref{ap:odd}, gauge
invariance, reflection symmetries and the fact that flow time parameter $t$  
has mass dimension $-2$ imply that non-trivial counterterms 
to the action must be even-dimensional.  In Eq.(\ref{eq:symact}) we have 
separated the effective action of the flow in the 4+1-dimensional volume, $S_{2,{\rm fl}}$, from
the action $S_{2,b}$ with support restricted to the 4-dimensional
boundary at $t=0$.  Both parts will be discussed in turn below.
Besides the effective action, also local observables are described by
an effective continuum field, again expanded in powers of $a^2$.
For a generic local observable ${\cal O}$ we write
\begin{equation}
   {\cal O}_{\text{eff}} = {\cal O}_0 + a^2 {\cal O}_2 + \mathcal O(a^4).
\label{eq:Oeff}
\end{equation}
To $\mathcal O(a^2)$ the Symanzik expansion of lattice expectation values then
takes the form, 
\begin{equation}
  \langle {\cal O} \rangle^\text{lat} = \langle {\cal O}_0 \rangle +
  a^2 \langle {\cal O}_2 \rangle 
  - a^2  \langle {\cal O}_0 S_{2,{\rm fl}} \rangle_c - a^2  \langle {\cal O}_0
  S_{2,b}\rangle_c +\mathcal O(a^4).
\end{equation}
Here, the expectation values on the RHS are defined in the continuum
theory with 
respect to the continuum action $S_0^{\text{cont}}$, 
and the notation $\langle \cdot\rangle_c$ serves as a reminder that only the
connected part contributes to the correlation functions with
counterterm insertions, for instance 
\begin{equation}
 \langle {\cal O}_0 S_{2,{\rm fl}} \rangle_c = \langle {\cal O}_0 S_{2,{\rm fl}} \rangle -
 \langle {\cal O}_0 \rangle \langle S_{2,{\rm fl}} \rangle\,.
\end{equation}

As the next step in the Symanzik procedure one determines a basis of
counterterms both for 
the action and the observables of interest. In the case of the action
these take the form 
\begin{eqnarray}
  S_{2,{\rm fl}}[B,L] &=& \int_0^\infty \int d^4x\,\sum_{i=1}^{n_{\rm fl}} Q_i(t,x), \\
  S_{2,b}[B,L] &=&  \int d^4x\,\sum_{i=1}^{n_b} O_i(x),
\end{eqnarray}
where the fields $Q_i(t,x)$ are gauge invariant polynomials in the
fundamental fields $B_\mu(t,x)$, $L_\mu(t,x)$ 
and their (space-time and/or flow time) derivatives, and the $O_i(x)$
are similarly constructed, but evaluated at $t=0$. Since $a^2
S_{2,{\rm fl}}$ must be dimensionless 
the fields $Q_i$ must have mass dimension~8 and otherwise share all
the symmetries with the lattice 
theory. The fields $O_i$ are dimension~6 fields, localized at the $t=0$ boundary. 
One of the important outcomes of the Symanzik analysis are the numbers
$n{\rm fl}$ and $n_b$ of basis elements, where fields differing by total (space-time) derivative terms
are considered equivalent. Furthermore, restricting to on-shell improvement the field
equations for $L_\mu$, $B_\mu$ 
and $A_\mu$ can be used to simplify the basis. 
Given  a basis of counterterms the final step of Symanzik's procedure
consists in adding 
lattice representatives of these operators to the lattice action, such
that, with appropriately chosen 
coefficients, the terms $S_{2,{\rm fl}}$ and $S_{2,b}$ are eliminated in
Symanzik's effective action for the improved lattice action.

A similar analysis then needs to be carried out for each observable
${\cal O}$ of interest, i.e.~${\cal O}_2$ in Eq.~(\ref{eq:Oeff}) is given
as a linear combination of local fields of mass dimension $\text{dim}({\cal O}_0)+2$ which
share all the lattice symmetries with ${\cal O}$.
While this procedure applies to any observables, we will
here focus on  gradient flow observables, i.e.~gauge invariant composite fields with
support at strictly positive flow times. 

If the full Symanzik procedure as outlined above were really
necessary, $\mathcal O(a^2)$ improvement 
would probably remain an academic curiosity. In particular, a rather
long list 
of dimension~8 counterterms for $S_{2,{\rm fl}}$ could be written
down, with little 
hope for practical relevance, so that one might be tempted to give up
on systematic $\mathcal O(a^2)$ improvement. 

Before proceeding along these lines, however, it is advisable to have
a closer 
look at this particular theory. As shown by L\"uscher and Weisz, the
theory is  perturbatively renormalizable to all orders in the
4-dimensional gauge 
coupling $g$~\cite{Luscher:2011bx}. More precisely, if one restricts
attention to gauge invariant observables, one just needs to renormalize 
the gauge coupling in the usual way, and also the quark masses if
the boundary  theory is generalized to QCD\footnote{We assume here
  that the quark 
  fields only live at $t=0$, i.e~they are not propagated into the
  4+1-dimensional bulk. For generalizations cf.~\cite{Luscher:2013cpa}} .
Moreover, any composite fields defined at finite flow-time are
automatically renormalized and do not mix with any other fields of the
same or lower canonical dimension. 
The action density (\ref{eq:actden}) is a typical example: its
renormalization at flow time $t=0$ requires the subtraction of both a quartic
and a logarithmic divergence.
None of this is required at finite $t$. It is instructive to consider
leading order perturbation theory to get a basic understanding of the
mechanism at work. 
Effectively, at finite flow time $t$, integrals over the loop-momentum
$p$ are cut 
off by an exponential suppression factor $\propto \exp(-2tp^2)$ in the
integrand. This 
renders most momentum integrals finite, so that one is only left with
those divergences which are cancelled by the standard counterterms in
the boundary theory. 

Hence the 4+1-dimensional theory enjoys rather special properties. In
particular, the field $L_\mu$ plays the r\^ole of a Lagrange
multiplier field 
which enforces the gradient flow equation as a constraint. The smoothening
properties of this equation are related to the fact that
perturbation theory only generates tree diagrams for
the correlation functions of gradient flow observables~\cite{Luscher:2011bx}.
The Symanzik expansion is then very much simplified
as we expect the following to hold:
\begin{itemize} 
\item The absence of bulk loop diagrams in the perturbative expansion of
  gradient flow observables implies that classical improvement
  of the flow action yields the $\mathcal O(a^2)$ effects {\em exactly},
  i.e.~without any corrections. 
\item By the same argument, non-perturbative $\mathcal O(a^2)$ improvement of
  cmoposite operators at positive flow time can be achieved
  by choosing discretizations that do not generate $\mathcal O(a^2)$
  effects when expanded classically.  
\item The only $\mathcal O(a^2)$ counterterms which receive genuine quantum
  corrections are the ones living in the 4-dimensional boundary at $t=0$.
  The full Symanzik procedure outlined above thus needs to be applied only to
  the $t=0$ boundary part, $S_{2,b}$ of the Symanzik action, and of course
  to any observable which is at least in part localised at the $t=0$
  boundary. 
\end{itemize}
In the following we first remind the reader of the classical
$a$-expansion and then address these points in the subsequent
subsections one at a time. 

\subsection{The classical $a$-expansion} 

According to the preceding discussion the counterterms appearing in
$S_{2,{\rm fl}}$ and 
in ${\cal O}_2$ for gradient flow observables are completely determined by
classically expanding the lattice action in the 4+1-dimensional volume
and the observables under consideration to order $a^2$.
The classical expansion assumes that the lattice approximates
an underlying continuum space-time manifold on which a smooth continuum
gauge field, $B_\mu(t,x)$, is defined. The lattice gauge field, $V_\mu(t,x)$,
is then related to the continuum gauge field by
parallel transport along the lattice links.
Parameterizing the path along the lattice link from $x+a\hat\mu$ to $x$
by $z(u) = x + (1-u)a\hat\mu$ (with parameter $u\in[0,1]$),
the precise relation is obtained by
iteratively solving the differential equation,
\begin{equation}
 \left\{\frac{d}{du} + B_\mu(t,z(u))\right\}v(u) = 0,\qquad v(0)=\unit.
\end{equation}
The solution, $v(u=1)\equiv V_\mu(t,x)$, can be concisely written in
terms of a path-ordered exponential,
\begin{eqnarray}
 V_\mu(t,x) &=& {\cal P} \exp\left\{ a \int_0^1 du\, B_\mu\left(t,z(u)\right)\right\}
          \label{eq:Vpathord}\\
          &=& \unit + a \int_0^1 du\,\, B_\mu\left(t,z(u)\hat\mu\right) \nonumber\\
          && \mbox{} + a^2 \int_0^1 du_1 \int_0^{u_1} du_2\,\,
              B_\mu\left(t,z(u_1)\right) B_\mu\left(t,z(u_2)\right) +\mathcal O(a^3)
           \label{eq:pexpexpansion}\\
          &=& \unit + aB_\mu(t,x) + \frac12 a^2 \left(\partial_\mu B_\mu(t,x) + B_\mu^2(t,x)\right)
          + \mathcal O(a^3).
\end{eqnarray}
While it is straightforward to carry out the expansion around $a=0$,
in practice, even a simple gauge invariant quantity like
the trace of the plaquette contains 4 link variables which need
to be expanded and combined to fourth order in $a$ to obtain the leading non-trivial term.
It is therefore highly advisable to perform the expansion efficiently
(cf.~e.g.~\cite{Luscher:1984xn,GarciaPerez:1993ki}).
We here follow L\"uscher and Weisz~\cite{Luscher:1984xn}, who, for fixed indices $\mu$ and $\nu$,
proposed to work in the following gauge:
\begin{equation}
  B_\mu(t,x) = 0 \quad\text{for all $x$}; \qquad
   B_\nu(x) = 0 \quad \text{if $x_\mu=0$}.
\label{eq:lwgauge}
\end{equation}
As a result, the expansion around $x=0$ is very much simplified.
For example, the plaquette field,
\begin{equation}
  P_{\mu\nu}(t,x) = V_\mu(t,x)V_\nu(t,x+a\hat\mu)V_\mu(t,x+a\hat\nu)^\dagger V_\nu(t,x)^\dagger,
  \label{eq:plaqp}
\end{equation}
is reduced to a single link,
\begin{equation}
  P_{\mu\nu}(t,0) = V_\nu(t,a\hat\mu) = {\cal P} \exp\left\{ a \int_0^1 du
     B_\nu\left(t,a\hat\mu + (1-u)a\hat\nu\right)\right\}.
\end{equation}
Recalling the definition of the path ordered exponential~(\ref{eq:pexpexpansion})
one needs the expansion of the $B$-field around $a=0$,
\begin{eqnarray}
  aB_\nu\left(t,a\hat\mu + \kappa a\hat\nu\right) &=& a^2\partial_\mu B_\nu(t,0)
  + \frac12 a^3\left\{\partial_\mu^2+ 2\kappa\partial\mu\partial_\nu\right\} B_\nu(t,0)\nonumber\\
  && + \frac16 a^4\left\{ \partial_\mu^3+3\kappa\partial_\mu^2\partial_\nu^{}
  + 2\kappa^2\partial_\mu^{}\partial_\nu^2\right\} B_\nu(t,0)\\
  && + \frac{1}{24} a^5 \left\{\partial_\mu^4 + 4\kappa \partial_\mu^3\partial_\nu^{}
      +6\kappa^2\partial_\mu^2\partial_\nu^2
      + 4 \kappa^3\partial_\mu^{}\partial_\nu^3\right\}B_\nu(t,0) +\dots,
      \nonumber
\end{eqnarray}
where $\kappa$ is a constant and neglected terms are of order $a^6$.
Following \cite{Luscher:1984xn} the gauge covariant expressions
can be {\em unambiguously} restored, with the result,
\begin{eqnarray}
  aB_\nu\left(t,a\hat\mu + \kappa a\hat\nu\right) &=& a^2 G_{\mu\nu}(t,0)
  + \frac12 a^3\left\{D_\mu + 2\kappa D_\nu\right\} G_{\mu\nu}(t,0) \nonumber\\
  && + \frac16 a^4\left\{ D_\mu^2 + 3\kappa D_\nu D_\mu  + 3\kappa^2 D_\nu^2\right\} G_{\mu\nu}(t,0)\\
  && + \frac{1}{24} a^5 \left\{D_\mu^3 + 4\kappa D_\nu^{} D_\mu^2
  + 6\kappa^2 D_\nu^2 D_\mu^{} + 4\kappa^3 D_\nu^3\right\}G_{\mu\nu}(t,0) +\ldots \nonumber
\end{eqnarray}
Inserting into the path ordered exponential with appropriate replacements for $\kappa$,
we thus obtain the gauge covariant expansion for the plaquette field,
\begin{eqnarray}
  P_{\mu\nu} &=& \unit + a^2G_{\mu\nu}+ \frac12 a^3 (D_\mu+D_\nu)G_{\mu\nu} \nonumber\\
  && \mbox{}+ \frac{1}{12}a^4\left\{\left(2D_\mu^2+3D_\nu^{} D_\mu^{} + 2D_\nu^2\right)G_{\mu\nu}
  +6\, G_{\mu\nu}G_{\mu\nu}\right\} \nonumber\\
  && \mbox{}+ \frac{1}{24}a^5\left\{ D_\mu^3+2D_\nu^{}D_\mu^2
     + 2 D_\nu^2 D_\mu^{} + D_\nu^3\right\}G_{\mu\nu}\nonumber\\
  && \mbox{}+ \frac{1}{12}a^5\left\{ (3 D_\mu+2 D_\nu)(G_{\mu\nu})^2
     + 2 G_{\mu\nu}D_\nu G_{\mu\nu}\right\} +\mathcal O(a^6),
\end{eqnarray}
which holds for any argument $(t,x)$. Similar expressions can be
derived for the other 3 plaquettes in the $\mu-\nu$ plane:
\begin{eqnarray}
  Q_{\mu\nu}(t,x) &=& V_\nu(t,x-a\hat\nu)^\dagger V_\mu(t,x-a\hat\nu) V_\nu(t,x+a\hat\mu-a\hat\nu)V_\mu(t,x)^\dagger,
   \label{eq:plaqq}\\
  R_{\mu\nu}(t,x) &=& V_\mu(t,x-a\hat\mu)^\dagger V_\nu(t,x-a\hat\mu-a\hat\nu)^\dagger
                    V_\mu(t,x-a\hat\mu-\hat\nu) V_\nu(t,x-a\hat\nu),
   \label{eq:plaqr}\\
  S_{\mu\nu}(t,x) &=& V_\nu(t,x)V_\mu(t,x-a\hat\mu+a\hat\nu)^\dagger V_\nu(t,x-a\hat\mu)^\dagger V_\mu(t,x-a\hat\mu),
     \label{eq:plaqs}
  \end{eqnarray}
and the next few orders can be obtained with moderate additional effort.

\subsection{Determination of $S_{2,{\rm fl}}$}

To find the bulk counterterm action $S_{2,{\rm fl}}$ we simply need to
apply the classical expansion to the bulk action in Eq.~(\ref{eq:act4plus1}).
This essentially amounts to the $a$-expansion of
the gradient flow equation, i.e.~$F_\mu(t,x)$ in Eq.~(\ref{eq:Fmudef}).
For the first term we find, in the L\"uscher-Weisz gauge (\ref{eq:lwgauge}),
\begin{equation}
 a^{-1} \left[\partial_tV_\mu(t,0)\right]V_\mu(t,0)^\dagger
= \int_0^1 du\, \partial_t B_\mu\left(t,(1-u)a\hat\mu\right),
\end{equation}
as all other terms are proportional to $B_\mu(t,0)=0$. The Taylor expansion can be easily performed to
all orders in $a$ with the result
\begin{equation}
   \int_0^1 du\, \partial_t B_\mu\left(t,(1-u)a\hat\mu\right)
= \sum_{n=0}^\infty \frac{a^n}{(n+1)!}\partial_\mu^n\partial_tB_\mu(t,0)\,.
\end{equation}
We therefore expect that the correct gauge covariant expression at any
lattice point $x$ must read 
\begin{equation}
  a^{-1} \left[\partial_tV_\mu(t,x)\right]V_\mu(t,x)^\dagger =
 \partial_tB_\mu(t,x) + \sum_{n=1}^\infty \frac{a^n}{(n+1)!}
 D_\mu^n\partial_tB_\mu(t,x)\,.
\label{eq:flow_LHS}
\end{equation}
At this point one may wonder whether the gauge covariant expression
really follows  unambiguously from the gauge fixed expansion,
in particular, whether the $t$-derivative always has to  be to the
right of the covariant 
$\mu$-derivatives.  That this is indeed correct can be established
by using the 4+1-dimensional gauge symmetry (cf.~Subsect~2), which implies
that the $a$-expansion of this term must be given as covariant
derivatives acting on $G_{4\mu}$.

Turning to the second term of (\ref{eq:Fmudef}), i.e.~the gradient
force term, 
we choose a quite general lattice gauge action parameterized by $c_{0,1,2}$
which includes all 4- and 6-link Wilson loops (plaquettes, rectangles,
chairs) 
except the twisted chairs/parallelograms. We decompose the action as
follows: 
\begin{equation}
  \label{eq:flact}
 S_{\text{g}}[V;c_0,c_1,c_2] = c_0  S_{\text{g,pl}}[V] + c_1  S_{\text{g,re}}[V]
    + c_2  S_{\text{g,ch}}[V].
\end{equation}
We first express the gradient force in terms of plaquettes and their covariant derivatives.
For the plaquette action we then find
\begin{equation}
 g_0^2\partial_{x,\mu} S_{\text{g,pl}}[V]
= \sum_\nu \left(P_{\mu\nu}(t,x)+Q_{\mu\nu}(t,x)^\dagger\right)_{\rm AH}\,,
\end{equation}
where we have introduced the projection on the trace-less
anti-hermitian part, i.e.~for an 
$N\times N$ matrix $M$ in colour space we define
\begin{equation}
  \left(M\right)_{\rm AH} = -2{\rm tr}\left(T^a M\right)\, T^a
\end{equation}
For the rectangle action we find,
\begin{eqnarray}
 g_0^2\partial_{x,\mu} S_{\text{g,re}}[V] &=&
  \sum_\nu\Bigl( 2 P_{\mu\nu}(t,x)P_{\mu\nu}(t,x) -  2 Q_{\mu\nu}(t,x)Q_{\mu\nu}(t,x) \nonumber\\
 &&\mbox{} + P_{\mu\nu}(t,x)S_{\mu\nu}(t,x) -   R_{\mu\nu}(t,x)Q_{\mu\nu}(t,x) \nonumber\\
 &&\mbox{} + \left(a\nabla_\mu P_{\mu\nu}(t,x)\right) P_{\mu\nu}(t,x)
    - Q_{\mu\nu}(t,x)a\nabla_\mu Q_{\mu\nu}(t,x) \nonumber\\
 &&\mbox{} + \left(a\nabla_\nu^\ast Q_{\mu\nu}(t,x)\right) Q_{\mu\nu}(t,x)
    + P_{\mu\nu}(t,x)a\nabla_\nu P_{\mu\nu}(t,x)\Bigr)_{\rm AH},
\end{eqnarray}
and a similar but slightly more complicated expression is obtained for the chairs.
Expanding each term to order $a^2$ and recombining them we get
\begin{eqnarray}
g_0^2\partial_{x,\mu} S_{\text{g}} &=&
    a^3\sum_{\nu} \Biggl\{ \left(c_0+8c_1+16c_2\right)\left(D_\nu G_{\nu\mu}
   + \frac{a}{2} D_\mu D_\nu G_{\nu\mu}\right) \nonumber\\
 &&\mbox{} + a^2\Biggl[\frac{1}{12}\left(c_0+20c_1+4c_2\right)
   \left(D_\nu^3+2D_\nu D_\mu^2\right) +(c_2-c_1)D_\mu^2 D_\nu \nonumber\\
 &&\hphantom{01234} + c_2\sum_{\rho}\left(3 D_\rho^2 D_\nu - 4D_\rho D_\nu D_\rho
    + 2 D_\nu D_\rho^2\right)\Biggr] G_{\nu\mu}\Biggr\}
   + \mathcal O(a^6),
\end{eqnarray}
where the arguments $(t,x)$ on the RHS have been omitted.
Collecting all results we define the expansion coefficients
\begin{equation}
  \label{eq:FmuTseries}
   F_\mu(t,x) = \sum_{n=0}^{\infty} a^n F_\mu^{(n)}(t,x),
\end{equation}
where the leading term defines the continuum limit,
\begin{equation}
   F_\mu^{(0)}(t,x) = \partial_t B_\mu(t,x) - (c_0+8c_1+16c_2) \sum_\nu D_\nu G_{\nu\mu}(t,x).
\end{equation}
Hence the correct normalization to reproduce the Yang-Mills gradient flow equation~(\ref{eq:YMflow}) is
$c_0+8c_1+16c_2=1$, which we use to eliminate $c_0$ in the higher order terms:
\begin{eqnarray}
  F_\mu^{(1)} &=& \frac12 D_\mu\left(\partial_t B_\mu - \sum_\nu D_\nu G_{\nu\mu}\right),\\
  F_\mu^{(2)} &=& \frac16 D_\mu^2 \partial_t B_\mu - \left(\frac{1}{12}
                + c_1-c_2\right)\sum_{\nu}\left(2 D_\nu^{}D_\mu^2+D_\nu^3\right) G_{\nu\mu}\nonumber\\
              && \mbox{} + \sum_{\nu}\left[(c_1-c_2) D_\mu^2D_\nu^{}
                 -c_2\sum_{\rho}\left(3D_\rho^2D_\nu^{}-4D_\rho^{} D_\nu^{} D_\rho^{}
                              +2 D_\nu^{}D_\rho^2\right)\right]G_{\nu\mu}.
\end{eqnarray}
Before proceeding we remark on the presence of odd powers of $a$ in
the expansion, which seems 
at odds with our expectation that only even powers of $a$ occur in
this theory. 
The resolution of this apparent contradiction lies in the fact that
the lattice fields 
$F_\mu(t,x)$ and $L_\mu(t,x)$ should be defined on the lattice link
connecting $x$ and $x+a\hat \mu$, rather than at the lattice site $x$.
In Appendix~B we demonstrate how the covariant re-expansion about the
midpoint 
of the link, $\tilde{x}  = x + \frac12 a\hat\mu$, eliminates such
terms. While this problem will not affect our discussion of the
$\mathcal O(a^2)$ counterterms, it clarifies that the corrections terms are
indeed of order $a^4$. 

We now proceed and work out the simplifations due to the field equations
for $B_\mu(t,x)$ and $L_\mu(t,x)$. Varying the continuum action with respect
to $L_\mu$ one obtains the Yang-Mills flow equation (\ref{eq:YMflow}),
whereas the variation with respect to $B_\mu(t,x)$ yields
\begin{equation}
\label{eq:eomB}
 \partial_t L_\mu = \sum_\nu \left( D_\mu  D_\nu L_\nu + D_\nu^2 L_\mu\right).
\end{equation}
Using the flow equation eliminates the O($a$) term $F_\mu^{(1)}$, and
this is the reason 
why the $\mathcal O(a^2)$ terms remain unaffected by the symmetrization about
the midpoint $\tilde{x}$, once the field
equations are taken into account. From the continuum flow equation we derive
\begin{equation}
   \partial_t \sum_{\nu}D_\nu G_{\nu\mu} =
   \sum_{\nu,\rho}\left(3D_\rho^2D_\nu^{}-4D_\rho^{} D_\nu^{}
   D_\rho^{} +2 D_\nu^{}D_\rho^2\right)G_{\nu\mu}.
\end{equation}
This allows to rewrite the $\mathcal O(a^2)$ term as follows:
\begin{eqnarray}
  F_\mu^{(2)}(t,x) &=&  \sum_\nu\Biggl\{
     -\left(\frac{1}{12} + c_1-c_2\right)\left(2
                       D_\nu^{}D_\mu^2+D_\nu^3\right) \nonumber\\
   && \hphantom{0123} +\left(\frac{1}{6} + c_1-c_2\right)D_\mu^2 D_\nu^{}
      -c_2\partial_t D_\nu^{}\Biggr\}G_{\nu\mu}(t,x).
\end{eqnarray}
From the corresponding $\mathcal O(a^2)$ flow action,
\begin{equation}
   S_{2,{\rm fl}}[B,L] = -2\int_0^\infty dt \int d^4x
   \sum_{\mu}\tr\left\{ L_\mu(t,x) F_\mu^{(2)}(t,x)\right\},
\end{equation}
one may now directly read off the counterterm structures $Q_i$ that
correspond  with a given choice of the coefficients $c_{1,2}$.
Unfortunately, there does not seem to be a choice such that $S_{2,{\rm fl}}$
vanishes. 
We also attempted to use Eq.~(\ref{eq:eomB}) as follows: considering the term
\begin{equation}
    2 c_2 \int_0^\infty dt \int d^4x\sum_{\mu,\nu}
    \tr\left\{ L_\mu(t,x) \partial_t D_\nu^{}G_{\nu\mu}(t,x) \right \},
\end{equation}
one may perform an integration by parts with respect to $t$. This
generates a surface term at $t=0$, 
\begin{equation}
    - 2 c_2 \int d^4x \sum_{\mu,\nu}
    \left.\tr\left\{ L_\mu(t,x) D_\nu^{}G_{\nu\mu}(t,x) \right
      \}\right\vert_{t=0} \,,
\end{equation}
which re-defines a coefficient of the counterterms entering $S_{2,b}$
(cf.~Subsect.~\ref{s2b}). 
Eq.~(\ref{eq:eomB}) then leads to space-time derivatives acting
on $L_\mu$, which can be integrated by parts (no surface terms are
generated here) 
to redefine $F_\mu^{(2)}$. Unfortunately, this does not yield a solution
with $S_{2,{\rm fl}} = 0$ either. We notice, however, that $S_{2,{\rm fl}}$ with the
L\"uscher-Weisz  choice of coefficients $c_1=-1/12$ and $c_2=0$, has a rather simple
structure, 
\begin{equation}
   \left.S_{2,{\rm fl}}\right\vert_{\rm LW} = -2\int_0^\infty dt \int d^4x
   \sum_{\mu,\nu}\tr\left\{L_\mu(t,x) \frac{1}{12} D_\mu^2D_\nu^{}
     G_{\nu\mu}(t,x)\right\}. 
\end{equation}
To cancel this term is relatively straightforward. Starting from the lattice
gradient force defined with the L\"uscher-Weisz action, $S_{\rm LW}$, we
simply act with,
\begin{equation}
  1+ \frac{1}{12} a^2 \nabla_\mu^\ast\nabla_\mu^{},
\end{equation}
on this gradient force, which yields the ``Zeuthen flow" equation
(\ref{eq:impflowlat}). 
The flow action $S_{2,{\rm fl}}$ for the Zeuthen flow does indeed vanish,
i.e.~we have successfully implemented $\mathcal O(a^2)$ improvement in the
4+1-dimensional bulk. 

\subsection{$\mathcal O(a^2)$ improvement of $E(t,x)$}
\label{sec:obs}

We here consider only the simplest observable, namely the action density $E(t,x)$ of Eq.~(\ref{eq:actden})
The two most popular lattice discretisations of $E(t,x)$ are
referred to as plaquette (pl) and clover (cl) definitions, respectively.
They are either obtained from the Wilson plaquette action or based on the so called clover
leaf definition of the field strength tensor,
\begin{equation}
  G^{\text{cl}}_{\mu\nu}(t,x)= \frac{1}{8a^2}\left(P_{\mu\nu}(t,x)+Q_{\mu\nu}(t,x)
  +R_{\mu\nu}(t,x)+S_{\mu\nu}(t,x) \right)_{\rm AH}\,,
  \label{eq:clover_Gmunu}
\end{equation}
which uses the 4 plaquettes (\ref{eq:plaqp}), (\ref{eq:plaqq}--\ref{eq:plaqs}) in the $\mu-\nu$ plane.
The plaquette and clover lattice versions of $E(t,x)$ are now given by:
\begin{eqnarray}
  E^{\text{pl}}(t,x) &=& -\frac12 a^{-4}\sum_{\mu,\nu}
  \left[\tr\left(P_{\mu\nu}(t,x) + P_{\mu\nu}(t,x)^\dagger\right) -2N\right]\,, \\
  E^{\text{cl}}(t,x) &=& -\frac12 \sum_{\mu,\nu}\tr\{G^{\text{cl}}_{\mu\nu}(t,x) G^{\text{cl}}_{\mu\nu}(t,x)\}\,.
\end{eqnarray}
Pushing the classical $a$-expansion of the plaquette $P_{\mu\nu}$ (\ref{eq:plaqp}) to O($a^6$)
one obtains
\begin{eqnarray}
 E^{\text{pl}}(t,x) &=&  E^{\text{cont}}(t,x) + \frac{1}{24} a^2\sum_{\mu,\nu}
   \left[\tr \left(D_\mu G_{\mu\nu}(t,x)\right)^2 +\tr \left(D_\nu G_{\mu\nu}(t,x)\right)^2\right] \nonumber\\
   && -\frac14 a \sum_{\mu,\nu}\left(\partial_\mu+\partial_\nu\right) \tr \left(G_{\mu\nu}(t,x)\right)^2\nonumber\\
   &&\mbox{}-\frac{1}{24} a^2 \sum_{\mu,\nu}
   \left(2\partial_\mu^2+2\partial_\nu^2 + 3\partial_\mu\partial_\nu\right)
   \tr \left(G_{\mu\nu}(t,x)\right)^2 + \mathcal O(a^3),
\end{eqnarray}
with the continuum limit $E^{\text{cont}}(t,x)$ given by Eq.(\ref{eq:actden}).
Proceeding in this way for all 4 plaquettes of the clover leaf we obtain the classical expansion
\begin{eqnarray}
 E^{\text{cl}}(t,x) &=&  E^{\text{cont}}(t,x) + \frac16 a^2\sum_{\mu,\nu}
   \left[\tr \left(D_\mu G_{\mu\nu}(t,x)\right)^2 +\tr \left(D_\nu G_{\mu\nu}(t,x)\right)^2\right] \nonumber\\
   && -\frac{1}{12} a^2 \sum_{\mu,\nu}\left(\partial_\mu^2+\partial_\nu^2\right)
   \tr \left(G_{\mu\nu}(t,x)\right)^2 + \mathcal O(a^4).
\end{eqnarray}
Several remarks are in order. First, the $a$-expansion of the plaquette yields contributions at every
order in $a$, whereas the symmetries of the clover definition imply only even powers of $a$.
The odd powers of $a$ could be eliminated by averaging over the 4 plaquettes of the clover leaf, which,
due to the trace operation, coincide with $\tr[P_{\mu\nu}(t,x)]$ for appropriately displaced arguments $x$.
Second, note the total derivative terms which may appear at any order in $a$.
Such terms do not contribute to the expectation value
$\langle E(t,x)\rangle$, provided that the chosen set-up is translation invariant. This would
e.g.~be the case in a finite volume with periodic or twisted periodic boundary conditions, and
thus in the  limit of infinite volume. However, translation invariance no longer holds
with either Dirichlet or Neumann conditions\footnote{Such boundary conditions are
often imposed in the Euclidean time direction, combined with periodic boundary conditions in
the spatial directions. In this case one may distinguish between the electric and magnetic
components of $E(t,x)$. In the latter, total derivatives only appear in the spatial directions
and thus do not contribute to the expectation value.} as required for
the Schr\"odinger functional~\cite{Fritzsch:2013je} or with open
boundary conditions~\cite{Luscher:2014kea}.
Similarly, when considering higher correlation functions such
as the 2-point correlator of two fields $E(t,x)$ total derivative terms cannot be ignored.
We will here focus on the translation invariant case and from now on consider such total derivative terms negligible.
This eliminates all the odd powers of $a$ in the expansion of $E^\text{pl}(t,x)$.
Hence, both discretizations are on equal footing and
counterterms ${\cal O}_{2}$ for $E^{\text{pl}}$ and $E^{\text{cl}}$ are now
easily identified as the $\mathcal O(a^2)$ coefficients in the classical expansion.
Given both $a$-expansions we observe that the $\mathcal O(a^2)$ terms
have the same structure, with the coefficients in the clover definition being larger by
a factor of 4. In any case we observe that the linear combination
\begin{equation}
  E^{\text{pl-cl}}(t,x) = \frac43 E^{\text{pl}}(t,x) -\frac13  E^{\text{cl}}(t,x)\,,
\label{eq:Eplcl}
\end{equation}
defines an $\mathcal O(a^2)$ improved observable for which ${\cal O}_2$ vanishes.
An alternative $\mathcal O(a^2)$ improved definition of $E(t,x)$ can be obtained
from the action density of a tree-level improved lattice action
such as the L\"uscher-Weisz action (Eq.(\ref{eq:latac}) with
$c_0=5/3$, $c_1=-1/12$ and $c_{2,3}=0$). Here again,
any ambiguity in the definition of a density from the action
amounts to total derivative terms, which we consider negligible in
the present context.

\subsection{Determination of $S_{2,b}$}
\label{s2b}
In this subsection we list the gauge invariant local
fields of dimension 6  which may appear in the boundary action $S_{2,b}$ of Symanzik's
effective action. 
Disregarding total derivative terms with respect to the space-time
coordinates $x$, we find the following list of 7 candidate counterterms, 
\begin{eqnarray}
O_1(x) &=&  \sum_{\mu,\nu}\tr\{[D_\mu F_{\mu\nu}(x)] D_\mu F_{\mu\nu}(x)\},
  \label{eq:defO1}\\
O_2(x) &=&  \sum_{\mu,\nu,\rho}\tr\{[D_\mu F_{\nu\rho}(x)] D_\mu F_{\nu\rho}(x)\},
   \label{eq:defO2}\\
O_3(x) &=&  \sum_{\mu,\nu,\rho}\tr\{[D_\mu F_{\mu\nu}(x)] D_\rho F_{\rho\nu}(x)\},
  \label{eq:defO3}\\
O_4(x) &=&  \sum_{\mu,\nu}\tr\{L_\mu(0,x) D_\nu F_{\nu\mu}(x)\},
  \label{eq:defO4}\\
O_5(x) &=&  \sum_{\mu}\tr\{L_\mu(0,x) L_\mu(0,x)\},\\
O_6(x) &=&  \sum_{\mu,\nu} \partial_t\tr\{G_{\mu\nu}(t,x)G_{\mu\nu}(t,x)\}\vert_{t=0},\\
O_7(x) &=&  \sum_{\mu}\tr\{L_\mu(t,x)\partial_t B_\mu(t,x)\}\vert_{t=0},
\end{eqnarray}
where $F_{\mu\nu}$ denotes the field strength tensor of the fundamental gauge field.

Again we apply the field equations. The Yang-Mills flow equation implies
\begin{equation}
   \partial_t G_{\mu\nu}(t,x)
    =\sum_\rho\left[D_\mu D_\rho G_{\rho\nu} - D_\nu D_\rho G_{\rho\mu}\right],
\end{equation}
so that, after taking into account the boundary condition $G_{\mu\nu}|_{t=0} = F_{\mu\nu}$,
we have
\begin{equation}
  O_6 + 4 O_3 = \text{total derivative}, \qquad O_7 = O_4.
\end{equation}
This eliminates $O_{6,7}$. The field equation (\ref{eq:eomB}) is not useful here.
However, a third field equation can be derived by varying the action at $t=0$
with respect to the fundamental gauge field $A_\mu(x)$.
Technically this is best done by discretising only
the flow time in the 4+1 dimensional continuum action and taking the limit of
continuous flow time in the end. The resulting field equation
is\footnote{While the continuum derivation may seem rather formal we note that a lattice
version of this equation can be derived directly from the $\varepsilon$-regularized 4+1-dimensional lattice
action by a variation with respect to the link field $U_\mu(x)$, followed by the limit $\varepsilon\rightarrow 0$.}
\begin{equation}
 \label{eq:eomA}
   \frac{1}{g^2} \sum_{\nu}D_\nu F_{\nu\mu}(x) = - L_\mu(0,x).  
\end{equation}
This equation leads to the relations
\begin{equation}
   O_5 = - \frac{1}{g^2} O_4, \qquad O_3 = - g^2 O_4
\end{equation}
Hence one may also eliminate $O_{3,5}$ in favour of $O_4$.

At this point it is useful to recall the situation in the
standard 4-dimensional theory~\cite{Luscher:1984xn}.
In fact there is a 1-parameter family of $\mathcal O(a^2)$ improved actions,
which, to tree-level, are parameterized by $x_p$ as follows:
\begin{equation}
  \label{eq:coefimp}
   c_0 = 5/3 -24 x_p,\qquad c_1=-1/12 + x_p, \qquad c_2=x_p, \qquad c_3=0.
\end{equation}
Expanding the action classically, the free parameter $x_p$ is seen to
multiply the counterterm $O_3$. 
The counterterm $O_3$ is thus redundant for the improvement of
standard observables. In principle one may thus tune
the coefficients~(\ref{eq:coefimp}) to achieve $\mathcal O(a^2)$
improvement of both standard 
and gradient flow observables. In practice however,
these coefficients define the gauge action used in the Monte-Carlo
simulation and  
the corresponding effective coefficient of $O_3$ should be regarded as fixed.
One therefore needs to find an alternative way to achive improvement,
and we chose to implement the counterterm $O_4$
(cf.~Subsect.~{\ref{subsect:summary3}}). 

Finally, we remark that the use of the field equation (\ref{eq:eomA})
in the counterterm basis 
holds for counterterm insertions only up to contact terms, namely whenever 
the counterterm argument coincides with the location of some field in
the correlation function under study.  
Such contact terms are thus absent for gradient flow observables
localized at strictly positive flow times.
However, we expect these relations to hold more generally, i.e. even
if some fields in the correlation functions 
are defined at zero flow time. In this case we expect that the contact
terms which make the difference  
are of the same form as the $\mathcal O(a^2)$ counterterms to the fields in the
correlation function  
and therefore just redefine  these counterterm coefficients. This
parallels the discussion in ref.~\cite{Luscher:1996sc} of on-shell
$\mathcal O(a)$ improvement in 
lattice QCD with Wilson quarks. 

\subsection{Summary of Section 3 and some practical considerations} 
\label{subsect:summary3}
Sect.~3 contains the main results of this paper and may appear rather 
technical. We therefore provide a short summary and comment on the practical
implementation of the lattice counterterm $O_4$.

There is a natural way of interpreting the gradient flow as a
4+1-dimensional local quantum field theory.
The flow time $t$ plays the role of the coordinate in the fifth dimension,
which only takes on non-negative values ($t\ge 0$). 
The dynamics of the theory in the bulk ($t>0$) is completely fixed by the 
deterministic flow equation. The classical nature of the theory 
for $t>0$ allows to implement the Symanzik improvement programme in a rather 
simple way: all $\mathcal O(a^2)$ cutoff effects produced by integrating the
flow equation can be eliminated via a suitable discretization of the
flow equation, 
which can be determined by the classical expansion to $\mathcal O(a^2)$.
Similar considerations allow to define discretized flow observables
that are free of $\mathcal O(a^2)$ lattice artefacts.
The only remaining $\mathcal O(a^2)$ effects are generated by
the action  at the boundary $t=0$, and are genuine quantum
effects. They correspond to the usual $\mathcal O(a^2)$ counterterms
(\ref{eq:defO1}--\ref{eq:defO3}) 
in the 4-dimensional action affecting all lattice observables.

To implement an $\mathcal O(a^2)$ improved lattice action one first
has to choose 
an $\mathcal O(a^2)$ improved 4-dimensional lattice gauge action which
amounts 
to choosing coefficients $c_{0-3}$ in Eq.~(\ref{eq:latac})
appropriately. It is  well-known how $\mathcal O(a^2)$ improvement can be
implemented at tree-level, 
and also  to order $g_0^2$ in the case of the pure gauge
theory~\cite{Luscher:1985zq}. 
In addition one needs to incorporate a lattice version of $O_4$
such as to cancel the insertion of $O_3$ on observables
without changing the coefficients $c_{0-3}$.

To achieve this we remind the reader that the 4+1-dimensional set-up is
used only for the theoretical analysis, whereas in practice one
integrates the gradient 
flow equation numerically and evaluates any observable such as
$E(t,x)$ along the flow. 
It turns out that the insertion of $O_4$ can be realized by a change in the
initial condition at $t=0$ for the gradient flow equation. Since in
this case $A_\mu(x)$ and $B_\mu(0,x)$ are not the same we need to fix
the integration variables in the 
4+1-dimensional field theory. We choose to integrate over the
fundamental gauge field $A_\mu(x)$ and the flow field $B_\mu(t,x)$ for
$t>0$. Therefore on the lattice we choose to integrate over $U_\mu(x)$
and $V_\mu(t,x)$ for $t>0$. A shift in the initial condition can be
implemented via
\begin{equation}
\label{eq:cb}
  V_\mu(t,x)\vert_{t=0} = e^{c_b g_0^2\partial_{x,\mu} S_\text{g}[U]} U_\mu(x)\,,
\end{equation}
where $c_b$ is the free improvement coefficient,
and $S_\text{g}[U]$ any 4-dimensional lattice action. In the 4+1-dimensional
formulation with $\varepsilon$-discretized flow time, the fields
$V_\mu(0,x)$ and $U_\mu(x)$ only enter in the terms
\begin{equation}
  S_\text{g}[U_\mu] - 2 a^4 \sum_{x,\mu} \tr \left\{ L_\mu(0,x)\left[
    a^{-1}\left(V_\mu(\varepsilon,x)V^\dagger_\mu(0,x) - 1\right) -
    \varepsilon X_\mu(0,x) 
  \right]\right\}\,,
\end{equation}
where $X_\mu(t,x)$ is, up to terms of O($\varepsilon$), the RHS~of the
flow equation. Now we can trade all references to $V_\mu(0,x)$ into 
$U_\mu(x)$, that is our path integral variable. Using
Eq.~\eqref{eq:cb} we can write  
\begin{equation}
  V_\mu(\varepsilon,x)V^\dagger_\mu(0,x) =
  V_\mu(\varepsilon,x)U^\dagger_\mu(x) - 
  c_bg^2_0 \partial_{x,\mu}S_{\text g}[U] + \dots
\end{equation}
where the dots represent higher order terms in the lattice spacing. 
Therefore the shift in the initial condition is equivalent (up to
higher order corrections in $a$) to the insertion of the counterterm
\begin{equation}
  2c_ba^6\sum_{x} \hat O_4(x) = -2c_b a^3\sum_{x,\mu} \tr \left\{
    L_\mu(0,x)\left(g_0^2\partial_{x,\mu} S_\text{g}[U] \right)
  \right\} \,.
\end{equation}
Renaming the variable $V_\mu(0,x)$ to $U_\mu(x)$ 
gets us back to the previous situation with standard boundary
conditions, $V_\mu(0,x)=U_\mu(x)$ 
except for the extra $\hat{O}_4$ term in the action. Hence we have
successfully traded the modified boundary conditions for the flow equation
for the $O_4$ term in the lattice action. In the next section
we will determine its coefficient $c_b$ at tree-level of perturbation theory.


\section{Perturbative analysis}

In this section we will study the Symanzik $\mathcal O(a^2)$
improvement of the gradient flow in perturbation theory. This will
allow us first to determine the improvement coefficient $c_b$ to tree-level.
Second, the study of the Zeuthen flow both in small volumes and for 
different observables will allow us to check explicitly that the use
of a tree-level improved action for the simulation together with the
tree level value\footnote{Note that all the improvement coefficients
  $c_i$ and $c_b$ have a perturbative expansion of the form $c(g_0^2)
  = c^{(0)} 
  + g_0^2c^{(1)} + \dots$. Since we are only concerned with tree level
improvement we will omit the superscript ${}^{(0)}$ in all improvement
coefficients. } of $c_b$, the Zeuthen flow and a classically 
improved definition of the observable yields expectation values that
are free of 
$\mathcal O(a^2)$ effects at tree level. As observables we choose
first $E(t,x)$. The 
contributions by the action, flow and observable to the cutoff effects of
$\langle E(t,x) \rangle $ at tree level have been computed
recently~\cite{Fodor:2014cpa,Ramos:2014kka}. Here we will show that
the $\mathcal O(a^2)$ tree level cutoff effects are absent not only in
infinite volume, but also in a finite volume with twisted periodic
boundary conditions, 
where the additional scale $L$ leads to more stringent tests. Second
we will consider the  
connected correlation function for $E(t,x)E(s,y)$ and show that
$\mathcal O(a^2)$ improvement 
by the flow is also obtained in this case.

\subsection{Gauge fixing}

In perturbation theory one parametrizes the links in a neighbourhood of
a classical configuration as follows
\begin{equation}
  U_\mu(x) =  \exp(ag_0A_\mu(x));\quad  V_\mu(t,x) =  \exp(ag_0B_\mu(t,x))\,.
  \label{eq:ABpert}
\end{equation}
Note that this standard convention implies a re-scaling of the fields,
\begin{equation}
  A_\mu \longrightarrow g_0 A_\mu,\qquad  B_\mu \longrightarrow g_0 B_\mu,
\end{equation}
compared to the preceding sections. In perturbation theory it
is convenient to use gauge symmetry to simplify explicit computations.
In the context of the gradient flow, gauge fixing is performed
by studying the generalized flow equation
\begin{equation}
    \partial_t B_\mu^{(\alpha)}(t,x) = D_\nu^{(\alpha)}
    G_{\nu\mu}^{(\alpha)}(t,x) +  
  \alpha D_\mu^{(\alpha)}\partial_\nu B_\nu^{(\alpha)}(t,x) \,,\qquad
   B_\mu^{(\alpha)}(0,x) = A_\mu(x)\,.
\end{equation}
The superscript ${(\alpha)}$ serves as a reminder that covariant
derivatives and field 
strength are made of the modified flow field $B_\mu^{(\alpha)}(t,x)$,
i.e.~the solution of the above equation. Note that the original flow 
equation is recovered by setting $\alpha=0$. The key observation is
that gauge invariant observables are independent of
$\alpha$~\cite{ZinnJustin:1987ux,Narayanan:2006rf,Luscher:2011bx}. In order 
to see this, one only has to check that the gauge transformation 
\begin{equation}
  \label{eq:gtrans}
  B_\mu = \Lambda B_\mu^{(\alpha)}\Lambda^{-1} + 
  \Lambda \partial_\mu
  \Lambda^{-1} \,,
\end{equation}
where 
\begin{equation}
  \partial_t\Lambda =
  \alpha \Lambda \partial_\mu B_\mu \,;\quad
  \Lambda\big|_{t=0} = 1\,,
\end{equation}
transforms a solution of the flow equation with arbitrary $\alpha$
into one with $\alpha=0$. 

On the lattice the procedure is completely analogous. We consider the
generalized flow equation
\begin{equation}
  \label{eq:flowlatmd}
  a^2\partial_t V_\mu^\Lambda(t,x) = g_0^2 \left\{ 
        -\big[ \partial_{x,\mu} S_\text{g}(V^\Lambda) \big]
        + a^2\nabla_\mu^{\Lambda}\big[\Lambda(t,x)^\dagger\partial_t \Lambda(t,x)\big]
                                           \right\} V_\mu^\Lambda(t,x) \,,
\end{equation}
or, for the case of the Zeuthen flow,
\begin{equation}
  a^2\partial_t{V}_\mu^\Lambda(t,x) = g_0^2 \left\{ -\left(1+\frac{a^2}{12}
      \nabla_\mu^{\Lambda*} \nabla_\mu^\Lambda \right)\partial_{x,\mu} S_\text{LW}(V^\Lambda)
    + a^2\nabla_\mu^\Lambda[\Lambda^\dagger(t,x)\partial_t \Lambda(t,x)]
\right\}V_\mu^\Lambda (t,x)\,, 
\end{equation}
with initial condition with $V_\mu^\Lambda(0,x) = U_\mu(x)$. One then easily verifies
that the gauge transformation 
\begin{equation}
  V_\mu(t,x) = \Lambda(t,x)V_\mu^\Lambda(t,x)\Lambda(t,x+\hat\mu)^\dagger\,,
\end{equation}
transforms a solution with an arbitrary function $\Lambda(t,x)$ into
one with $\Lambda=1$. A natural choice for the function $\Lambda(t,x)$
then is given as the solution of the equation,
\begin{equation}
  \label{eq:lam}
  \Lambda^{-1}\partial_t \Lambda = \alpha
   \partial^\ast_\mu B_\mu(t,x)\,,\qquad
  \Lambda\big|_{t=0} = 1\,.
\end{equation}
Note that this is a particular application of the 4+1-dimensional gauge transformations
described in Subsect.~\ref{subsect:gaugesym} and it is thus clear that
gauge invariant observables remain unaffected by the choice of $\alpha$.
This can be turned around to provide checks on the correctness of a given
calculation. In the following we drop the indices $(\alpha)$ (or $\Lambda$) from the fields
and we will quote any intermediate results in Feynman gauge ($\alpha=1$).
Some elements used for our checks of gauge parameter independence are given in Appendix~\ref{ap:gauge}.

\subsection{Determination of $c_b$ to tree level}

We first assume that the lattice is infinitely extended
and expand the general class of actions, Eq.~(\ref{eq:latac}),
to leading order in the coupling\footnote{See
Appendix~\ref{ap:conv} for a summary of our notation and conventions.},
\begin{equation}
  S_\text{g}[U;\{c_i^{(a)}\}] = \frac12 \sum_{\mu,\nu}\int_p \tilde A_\mu^a(-p) K_{\mu\nu}^{(a)}(p;\lambda)
  \tilde A_\nu^a(p)  + \mathcal O(g_0)\,, 
\end{equation}
where $\lambda$ is a gauge fixing parameter and explicit expressions for the lattice kernels,
$K_{\mu\nu}^{(a)}(p;\lambda)$, are given in Appendix~\ref{ap:conv}.

Similarly, the flow equation contains the gradient of a lattice action
which, to leading order in the coupling, is parameterized by
another action kernel, $K^{(f)}_{\mu\nu}(p;\alpha)$. The flow equation to this
order then takes the form of the heat equation,
\begin{equation}
  \partial_t {\tilde B_\mu^a(t,p)} = -\sum_\nu K_{\mu\nu}^{(f)}(p;\alpha)
  \tilde B_\nu^a(t,p)\,.
  \label{eq:linflow}
\end{equation}
The initial condition for the flow equation Eq.~(\ref{eq:cb})
reads to leading order
in the fields\footnote{Note that higher orders in the fields imply
  higher order in the coupling, 
too, cf.~Eq.~(\ref{eq:ABpert}).},
\begin{equation}
  \tilde B_\mu(0,p) = \sum_\nu \left[\delta_{\mu\nu} + a^2c_b
    K_{\mu\nu}^{(i)}(p;0) \right] \tilde A_\nu(p)\,,
\end{equation}
where $K_{\mu\nu}^{(i)}$ is yet another action kernel. No gauge fixing
term is required here, so that the gauge parameter is set to zero.
The linearized flow equation (\ref{eq:linflow}) can now be solved easily
\begin{equation}
  \tilde B_\mu^a(t,p) = \sum_{\nu,\rho} H_{\mu\nu}(t,p;\alpha) \left[
    \delta_{\nu\rho} + a^2c_b
    K_{\nu\rho}^{(i)}(p;0) \right] \tilde A_\rho^a(p)\,,
    \label{eq:relBA}
\end{equation}
where $H_{\mu\nu}$ is the heat kernel given by
\begin{equation}
  H_{\mu\nu}(t,p;\alpha) = \exp\left(-t
    K^{(f)}(p;\alpha)\right)_{\mu \nu} \,. 
\end{equation}
Note that we have used here $K^{(f)}(p;\alpha)$ as a matrix with respect to the
Lorentz indices and the exponential has to be taken of that matrix.
In the following we will often make  use of such a matrix notation, in order to
avoid an abundance of Lorentz indices.

Finally, the observable $E(t,x)$, being an
action density, can be parameterized by a further lattice action kernel,
$K^{(o)}(p,0)$, with gauge fixing parameter set to zero.
To this order we then obtain for the expectation value
\begin{equation}
  \langle E(t,x) \rangle =
  \frac{N^2-1}{2} g_0^2  \int_p {\rm Tr}\left\{
  K^{(o)}(p;0)\,\bar{D}(t,t,p;\alpha,\lambda)\,\right\} + \mathcal O(g_0^4),
\end{equation}
where the trace is over Lorentz indices only and 
the gauge field propagator at positive flow time is defined by
\begin{equation}
  \langle \tilde B_{\mu}^a(s,p)\tilde B_\nu^b(t,q) \rangle
  = (2\pi)^4 \delta^{(4)} (p+q)\delta^{ab}\bar{D}_{\mu\nu}(p,s,t;\alpha,\lambda)  \,.
\end{equation}
Due to the relation~(\ref{eq:relBA}), this propagator depends implicitly on both
gauge parameters, $\alpha$ and $\lambda$, of the
flow equation and of the the action, respectively.
Introducing the standard 4-dimensional gauge field propagator
 \begin{equation}
  \langle \tilde A_{\mu}^a(p)\tilde A_\nu^b(q) \rangle
  = (2\pi)^4 \delta^{(4)} (p+q)\delta^{ab} D_{\mu\nu}(p;\lambda) \,,
\end{equation}
this propagator is the matrix inverse of the action kernel,
\begin{equation}
   K^{(a)}(p,\lambda) D(p,\lambda) = \unit,
\end{equation}
and the gauge fixing parameter $\lambda$ must be non-zero for the inverse to exist.
Using these ingredients, the gauge field propagator at positive flow
time can now be written as follows,
\begin{equation}
\begin{split}
  \bar{D}(p,s,t;\alpha,\lambda) = &
  H(s,p;\alpha)\left[\unit+a^2 c_b K^{(i)}(p;0)\right] \\
  & \times D^{(a)}(p,\lambda)
  \left[\unit+a^2 c_b K^{(i)}(-p;0)\right]^T H(t,-p;\alpha)^T
\end{split}
\end{equation}
where we have  denoted the matrix transpose by the superscript $T$.

In summary, the choices of action, flow and observable discretization
correspond to the choice of three action kernels. Finally the shift in
the initial condition is encoded in a fourth choice of
kernel. Explicit expressions for some popular choices of kernels
are given in the Appendix~\ref{ap:gauge}.

In order to obtain the leading order cutoff effects
we now expand the kernels as follows.
\begin{equation}
  K(p;\lambda) = K^{\rm cont}(p;\lambda) +
  a^2R(p;\lambda) +
  \mathcal O(a^4)\,,
\end{equation}
where the continuum kernel is given by
\begin{equation}
  K_{\mu\nu}^{\rm cont}(p;\lambda)  = p^2\delta_{\mu\nu} -
  (1-\lambda)p_\mu p_\nu\,.
\end{equation}
Using the continuum kernel only and neglecting cutoff effects we
thus obtain the well-known continuum result in infinite volume,
\begin{equation}
  \langle E(t,x) \rangle = g_0^2\mathcal E_0^{\rm cont}(t) + \mathcal O(g_0^4,a^2),\qquad
  \mathcal E_0^{\rm cont}(t)  = \frac{3(N^2-1)}{128 \pi^2 t^2} .
 \label{eq:Eoftcont}
\end{equation}
Explicit expressions for the correction terms $R_{\mu\nu}(p;\lambda)$
are given in Appendix~\ref{ap:gauge}. In order to compute the leading correction to the propagator
$D_{\mu\nu}(p;\lambda)$ and to the heat kernel $H_{\mu\nu}(t,p;\alpha)$ it is convenient s
to work in Feynman gauge ($\lambda=\alpha=1$), since in this case
$K_{\mu\nu}^{\rm cont}(p;1)$ is proportional to
$\delta_{\mu\nu}$. Working in a general gauge is however not much more difficult and
serves as a check that the gauge dependence actually cancels in the
final evaluation of the observable. A few technical details pertaining
to such a check are given in Appendix~\ref{ap:gauge}.

In the following we will use Feynman gauge
and remove the gauge parameters as arguments of the action and flow kernels.
We will also omit them in the kernels for the observable and initial conditions
however, with the understanding that they must be set to zero in these cases.
In Feynman gauge ($\lambda=\alpha=1$) it is straightforward to check that
\begin{subequations}
\label{eq:exp}
\begin{eqnarray}
  D_{\mu\nu}(p) &=& \frac{1}{p^2}\left[
\delta_{\mu\nu} - \frac{a^2}{p^2}R_{\mu\nu}(p)
\right]  + \mathcal O(a^4)\\
  H_{\mu\nu}(t,p) &=& e^{-tp^2}\left[
  \delta_{\mu\nu} - a^2tR_{\mu\nu}(p)
 \right] + \mathcal O(a^4)\,,
\end{eqnarray}
\end{subequations}
and finally, putting all the pieces together 
and after some algebra, we get
\begin{eqnarray}
  \nonumber
  \mathcal E_0(t) = \mathcal E_0^{\rm cont}(t)\left\{
  1 + \frac{a^2}{t}\Big[ \right. &&(d_1^{(o)}-d^{(a)}_1)J_{4,-2}
  + (d_2^{(o)}-d^{(a)}_2+2c_b)J_{2,0}
  - \\
  &&- \left. 2 d_1^{(f)}J_{4,0} -2 d_2^{(f)}J_{2,2}
  \Big]\right\} + \mathcal O(a^4)\,,
  \label{eq:Eoftinfvol}
\end{eqnarray}
where the constants $J_{n,m}$ are defined by
\begin{equation}
  J_{n,m} =  t^{(n+m)/2}\frac{\int_p e^{-2tp^2}\,
    (p^n)(p^m)}{\int_p e^{-2tp^2}}\,,
\end{equation}
and
\begin{equation}
  \label{eq:ratios}
  p^n = \left\{ 
    \begin{array}{ll}
      \sum_\mu (p_\mu)^n & n>0 \\
      \left[\sum_\mu (p_\mu)^n\right]^{-1} & n<0 \\
    \end{array}
  \right.\,.
\end{equation}
In fact it is straightforward to evaluate the integrals with the result,
\begin{equation}
  J_{4,-2} = 1/2,\quad
  J_{2,0}  = 1,\quad
  J_{4,0}  = 3/4,\quad
  J_{2,2}  = 3/2\,.
  \label{eq:Jratios}
\end{equation}
The coefficients $d_{1,2}^{(a,o,f)}$ must be
independent of the gauge parameters $\alpha$ and $\lambda$ and we have checked this explicitly.
Their values depend on the choices made for the various kernels.
For example, for a general action of the form
Eq.~(\ref{eq:latac}) we have 
\begin{subequations}
\label{eq:dval}
\begin{eqnarray}
d_1 &=& -\frac{1}{12}-\frac{2}{3}c_1+\frac{2}{3}c_2+\frac{2}{3}c_3\,,\\
d_2 &=& -\frac{1}{3}c_1 -\frac{2}{3}c_2-\frac{2}{3}c_3\,.
\end{eqnarray}
\end{subequations}
Table~\ref{tab:dcoef} summarizes the values of the coefficients
$d_{1,2}^{(a,o,f)}$ for the most common choices. It is easy to see
that the use of the Zeuthen flow together with the tree-level improved
L\"uscher-Weisz action and any classically improved discretization for
the observable (see section~\ref{sec:obs}) has no tree-level
$\mathcal O(a^2)$ cutoff effects as long as $c_b=0$. Therefore, to
tree-level, the L\"uscher-Weisz action ($c_1=-1/12, c_2=0$) produces
tree-level improved results for gradient flow observables. For the
case of a generalized tree-level improved action
Eq.~(\ref{eq:coefimp}) we have to choose  
\begin{equation}
  c_b = -\frac{1}{2}x_p,
\end{equation}
in order to obtain tree-level improvement.

\begin{table}
  \centering
  \begin{tabular}{l|ll}
  \textbf{Discretization} & $d_1$ & $d_2$\\
\hline
  Plaquette & $-1/12$ & 0 \\
  L\"uscher-Weisz & $-1/36$ & $1/36$ \\
  $\frac{4}{3}$ Plaquette $-\frac{1}{3}$ Clover & $-1/36$ & $1/36$ \\
  One-parameter tree-level improved & $-1/36$ & $1/36-x_p$ \\
  Clover & $-1/4$ & $-1/12$ \\
  Zeuthen & 0 & 0 \\
 \hline
  \end{tabular}
  \caption{Values of the coefficients in the $\mathcal O(a^2)$ terms of
    $t^2\langle E(t,x) \rangle$ in infinite volume. The one-parameter
    family of tree-level improved actions corresponds to the
    choice of coefficients Eq.~(\ref{eq:coefimp}), the L\"uscher-Weisz
  tree level improved action being the particular choice with $x_p=0$.}
  \label{tab:dcoef}
\end{table}

As the reader can see, besides the Zeuthen flow there seem to be many
ways to cancel  the tree-level $\mathcal O(a^2)$ effects (see
also~\cite{Fodor:2014cpa}), as these are encoded in a single term, once the
numerical values (\ref{eq:Jratios}) and for $d_{1,2}$
(cf.~Table~\ref{tab:dcoef})  are inserted into Eq.~(\ref{eq:Eoftinfvol}).
We are thus led to look for more stringent tests of $\mathcal O(a^2)$
improvement by looking  at a variety of observables and/or kinematics. After all,
rather than improving a particular observable in a specific situation
(e.g.~in infinite volume), Symanzik improvement is designed to work for any
observable in both finite and infinite volume.

\subsection{Twisted periodic boundary conditions}

A stringent test of our computations can be made when studying
$t^2\langle E(t,x) \rangle$ in a finite volume.
Due to the presence of a new scale $L$, the cutoff effects will in general depend on the
dimensionless ratio $c=\sqrt{8t}/L$. Improvement requires that the
tree-level cutoff effects vanish \emph{for all values of
  $c=\sqrt{8t}/L$}. 

As a finite volume renormalization scheme, we will use twisted
boundary conditions for our gauge field. In this setup, the gauge
field changes by a gauge transformation when displaced by a
period. Gauge invariant quantities are still periodic, but the absence
of zero-modes in the perturbative expansion turn out to 
be very convenient for our analytic computations. The
gradient flow has already been studied in this setup, and we will not
give much details here but refer the interested reader to the
work~\cite{Ramos:2014kla} and the references cited therein.  

We will only need the perturbative expression of $\langle
E(t,x)\rangle$ to leading order, given by
  
\begin{subequations}
  \label{eq:mastertw}
  \begin{equation}
  \langle E(t,x)\rangle = g_0^2\mathcal E_0(t,c) + \mathcal O(g_0^4)
\end{equation}
with
\begin{eqnarray}
  \nonumber
  \mathcal E_0(t,c) = \frac{c^4}{128t^2}
  \sideset{}{'}\sum_{P} {\rm Tr}&&\Big\{
  H^{(f)}(t,P)\left[
  \unit + a^2c_b 
  K^{(i)}(P) \right]D^{(a)}(P)
  \\
  \times &&  \left[
  \unit + a^2c_b 
  K^{(i)}(-P) \right]^T  H^{(f)}(t,-P)^T K^{(o)}(P)
  \Big\} \,. 
\end{eqnarray}
\end{subequations}
Note that the expression is almost identical to the
infinite volume one, except that the momentum integral has been substituted by
a sum (hardly a surprise). The particularities of the twisted boundary conditions are hidden in the
sum and momentum symbols. First notice that the momentum (with capital
letters $P_\mu$) can be uniquely decomposed as
\begin{equation}
  P_\mu = \frac{2\pi n_\mu}{L} + \frac{2\pi \tilde n_\mu}{NL},
\end{equation}
with $n_\mu = 0,\dots,L/a-1$ and
\begin{equation}
  \label{eq:ptw}
    \tilde n_\mu = \left\{\begin{array}{ll}
                            0, & \text{ if } \mu = 0,3\,, \\
                            0,\dots,N-1, & \text{ if } \mu = 1,2\,, 
\end{array}\right.
\end{equation}
i.e.~there is the usual space-momentum, but in the directions of the
twisted plane $x_1-x_2$ the momentum $P_\mu$ lives in an apparently
larger lattice of size $NL$, where $N$ is the rank of the gauge
group. Finally the sum symbol $\sideset{}{'}\sum_P$ means sum both
over $n_\mu$ and $\tilde 
n_\mu$, but without the terms with $\tilde n_1 = \tilde n_2 = 0$. In
particular the sum has no term with a zero total momentum. Notice that
the colour factor $N^2-1$ is produced by the sum over $\tilde n_\mu$. 

The algebra is very similar to the one of the previous section, with
the important difference that now the sums actually depend on the
dimensionless ratio $c=\sqrt{8t}/L$. In fact fixing the flow time in
units of the volume in this way we get
\begin{eqnarray}
\label{eq:twc}
  \nonumber
  \mathcal E_0(t,c) = \mathcal E_0^{\rm
  cont}(t,c) \bigg\{
  1 + \frac{a^2}{t^2}\Big[ &&
                                  (d_1^{(o)}-d^{(a)}_1)\mathcal J_{4,-2}(c) 
  + (d_2^{(o)}-d^{(a)}_2+2c_b)\mathcal J_{2,0}(c) - \\
                               &&- 2 d_1^{(f)}\mathcal J_{4,0}(c) -2 d_2^{(f)}\mathcal J_{2,2}(c)
     \Big]\bigg\}+\mathcal O(a^4)\,,
\end{eqnarray}
where
\begin{equation}
  \mathcal E_0^{\rm cont}(t,c) = \frac{3c^4}{128t^2} \vartheta_3^2(0|\imath \pi
  c^2) \left[ 
    \vartheta_3^2\left(0|\imath \pi
  c^2/N^2\right) - \vartheta_3^2(0|\imath \pi
  c^2)
  \right]
  \,,
\end{equation}
and the third Jacobi theta function reads 
\begin{equation}
  \vartheta_3\left(z|\tau\right) = \sum_n e^{\imath \pi \tau n^2} 
  e^{2\imath n z}\,.
\end{equation}
Finally the functions $\mathcal J_{i,j}(c)$ are given
by\footnote{Negative powers ($j<0$) have to be understood as in the infinite
  volume, Eq.~(\ref{eq:ratios}).}
\begin{equation}
  \label{eq:Jvfv}
  \mathcal J_{i,j}(c) =
  \left(\frac{c\pi}{\sqrt{2}}\right)^{i+j}
  \frac{\sideset{}{'}\sum_n \exp\{-c^2\pi^2(n + \tilde n/N)^2\}\,
  (n+\tilde n/N)^i (n+\tilde n/N)^j }{\vartheta_3^2(0|\imath \pi
  c^2) \left[ 
    \vartheta_3^2\left(0|\imath \pi
  c^2/N^2\right) - \vartheta_3^2(0|\imath \pi
  c^2)
  \right]}\,. 
\end{equation}

\begin{figure}
  \centering
  \includegraphics[width=0.8\textwidth]{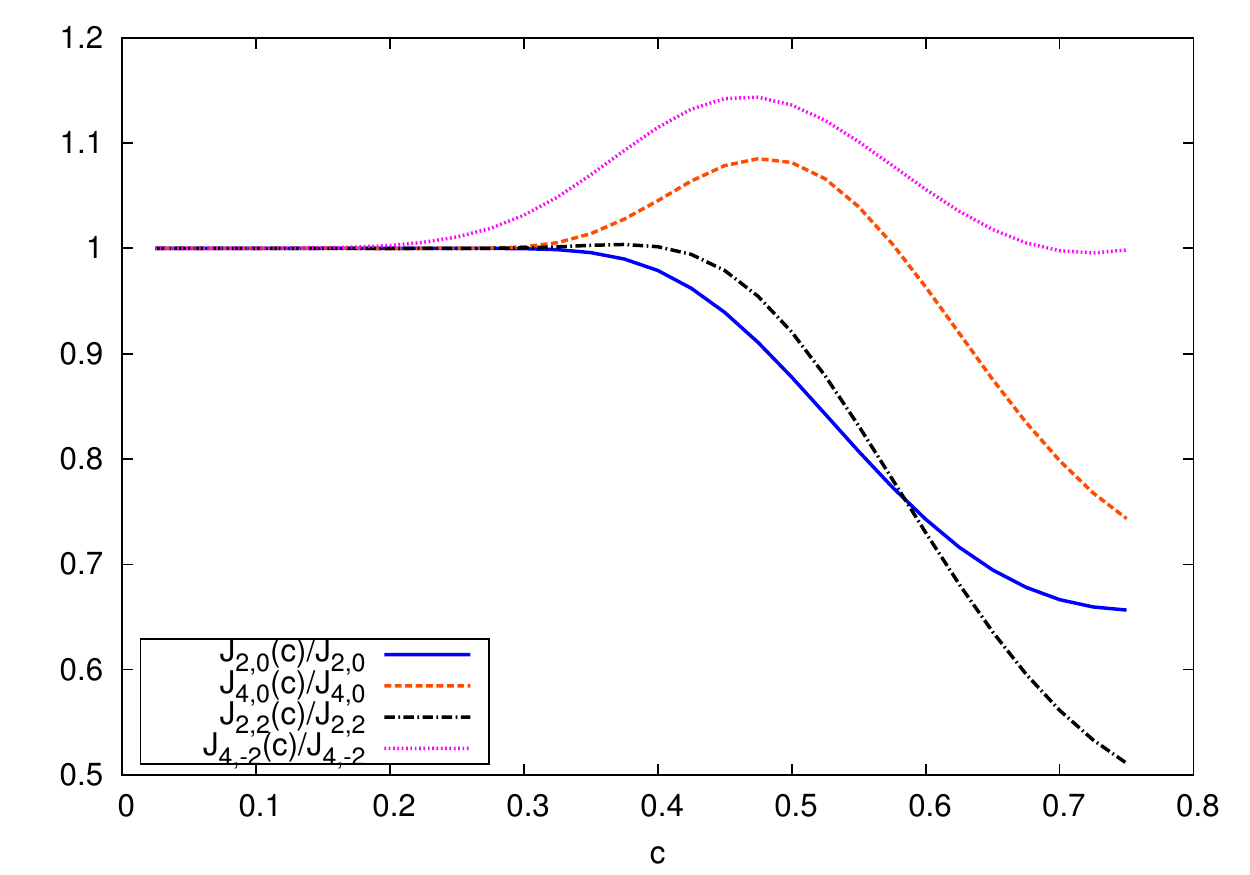}
  \caption{Ratio of the finite volume functions $\mathcal J_{i,j}(c)$
    (Eq.~\eqref{eq:Jratios}) 
    over the infinite volume predictions $J_{i,j}$
    (Eq.~\eqref{eq:Jvfv}). When $c>0.2$ there
    are significant differences between them. Moreover the different
    functions $\mathcal J_{i,j}(c)$ are in general linearly
    independent.}
  \label{fig:Jfunc}
\end{figure}

In the limit $c=\sqrt{8t}/L\rightarrow 0$, we recover the expressions of the
infinite volume, in particular
\begin{equation}
  \lim_{c\rightarrow 0} \mathcal J_{i,j}(c) = J_{i,j}\,,
\end{equation}
but for non-zero $c$ the functions $\mathcal J_{i,j}(c)$ are
in general linearly independent (see Fig.~\ref{fig:Jfunc}). The
coefficients $d_{1,2}^{(a,o,f)}$ 
are still the same, and the reader can check that the tree-level
$\mathcal O(a^2)$ cutoff effects given by expression Eq.~(\ref{eq:twc})
vanish for all values of $c$ when one uses our improved setup
(i.e. L\"uscher-Weisz action, Zeuthen flow and L\"uscher-Weisz
observable). Any other choice of improved action toghether with the
appropriate choice of $c_b$ also does the work. For this to happen it is
crucial that the flow coefficients 
$d_{1,2}^{(f)}$ are both zero, since the functions $\mathcal J_{4,0}(c)$ and
$\mathcal J_{2,2}(c)$ are linearly independent. In
particular it is easy now to check that the so called Symanzik flow in
the literature~\cite{Borsanyi2012} or any set of coefficients
in~\cite{Fodor:2014cpa}, does {\em not} remove the tree-level cutoff
effects in finite volume. 
For the Zeuthen flow both coefficients do identically vanish, so that
$\mathcal O(a^2)$ effects are indeed removed as expected
on theoretical grounds.

\subsection{The connected 2-point function of $E(t,x)$}

Further tests of the Zeuthen flow can be obtained by
considering different operators at positive flow time. In particular,
we now consider the 2-point function of $E(t,x)$ in a
periodic box of size $L$ with twisted periodic boundary
conditions (i.e.~the same setup as above),
\begin{equation}
  t^2 s^2\langle E(t,x)E(s,0)\rangle_c = t^2s^2\left[ \langle
    E(t,x)E(s,0)\rangle - \langle E(t,x)\rangle \langle
    E(s,0)\rangle\right]\,. 
\end{equation}
The factor $t^2s^2$ renders this quantity dimensionless, so that
it can be considered a function of the two dimensionless parameters,
\begin{equation}
  c = \frac{\sqrt{8t}}{L};\quad  d = \frac{\sqrt{8s}}{L}\,.
\end{equation}
Computing to leading order in the coupling
the result can be written in the form,
\begin{equation}
  t^2 s^2\langle E(t,x)E(s,0)\rangle_c = t^2 s^2g_0^4\mathcal M(t,s;x)
  + \mathcal O(g_0^6)\,,
\end{equation}
with,
\begin{eqnarray}
  \nonumber
  t^2 s^2\mathcal M(t,s;x) =
  \frac{c^4 d^4}{1024} &&\sum_{P, Q} e^{\imath (P+Q)x}
  \,{\rm Tr}\bigg\{
  K(P,Q) H(P,t) D(P) H(-P,s)^T\\ 
  &&  \times K(Q,P) H(-Q,t)^T D(Q)^T H(Q,s)
  \bigg\}\,.
\end{eqnarray}
The generalized kernel $K(P,Q)$ encodes the discretization of the observable.
Up to terms of $\mathcal O(a^2)$ it is given by 
\begin{equation}
  K_{\mu\nu}(P,Q) = K_{\mu\nu}^{\rm cont}(P,Q)+  \mathcal O(a^2),
\end{equation}
with the continuum kernel given by
\begin{eqnarray}
  K_{\mu\nu}^{\rm cont}(P,Q) &=& \sum_{\rho}P_\rho Q_\rho \delta_{\mu\nu} -
  P_\mu Q_\nu\,. 
\end{eqnarray}
The finite volume calculation for $\langle E(t,x)\rangle$ has taught us
that the $\mathcal O(a^2)$ contributions of the flow have to cancel by themselves,
i.e.~a cancellation with other $\mathcal O(a^2)$ contributions by the action or the
observable are not possible, due to the linear independence of the momentum sums.
In order to assess the improvement of the Zeuthen flow it is therefore enough
to focus on these $\mathcal O(a^2)$ contributions.
Using again the Feynman gauge for flow and action, we obtain
$\mathcal O(a^2)$ terms from the flow of the form,
\begin{equation}
\begin{split}
 -a^2\frac{c^4 d^4}{1024} &\sum_{P, Q}
  e^{\imath (P+Q)x} e^{-(t+s)\left(P^2+Q^2\right)} \frac{1}{P^2 Q^2} \\
  &\times \,{\rm Tr}\bigg\{
  K^{\rm cont}(P,Q) \left(tR(P)+sR(P)^T\right) K^{\rm cont}(Q,P)\bigg\}\,,
\end{split}
\end{equation}
and a second term with a similar structure. In both cases it is
useful to note the property of the kernel,
\begin{equation}
   K^{\rm cont}(P,Q) = T(Q) K^{\rm cont}(P,Q) T(P),
\end{equation}
where $T(P)$ is the transverse projector,
\begin{equation}
  T_{\mu\nu}(P) = \delta_{\mu\nu} - \frac{P_\mu P_\nu}{P^2}\,. 
\end{equation}
The $\mathcal O(a^2)$ correction to the Zeuthen flow kernel, $R^Z(P)$,
has the nice property that 
\begin{equation}
  T(P) R^Z(P) T(P) = 0.
\end{equation}
Hence we can conclude that the Zeuthen flow
does not contribute any $\mathcal O(a^2)$ effects to
this 2-point function either.
Due to the different Lorentz index structure of this case compared to
the simpler case of $E(t,x)$, and to the fact that now, in general,
the cutoff effects are functions of two variables $(c,d)$, this test
imposes further constraints on the possible improvement solutions. In
particular, the so called \emph{chair flow} in~\cite{Ramos:2014kka}, which
happens to also cancel the $\mathcal O(a^2)$ effects of $\langle E(t,x)\rangle$
\emph{in a finite volume}, can be shown to produce
$\mathcal O(a^2)$ contributions to the 2-point function considered here.


\section{Conclusions and Outlook}

We have systematically investigated the
structure of $\mathcal O(a^2)$ effects in flow quantities using
Symanzik's approach applied to the 4+1-dimensional local
formulation of the theory. Improvement to $\mathcal O(a^2)$ for
gradient flow quantities appears to be easier than 
one might have thought, mainly due to the classical nature of the
gradient flow equation. In particular the classical $a$-expansion is
sufficient to obtain the counterterms for both local composite operators at
positive flow time and the action in the 4+1-dimensional bulk
(i.e. due to the absence of loops in the bulk, no new counterterms are
generated). 

Our main results are summarized in the Zeuthen flow
equation~(\ref{eq:impflowlat}) and the  
improved lattice definitions of the observable $E(t,x)$, 
either as linear 
combination of clover and plaquette definitions~(\ref{eq:Eplcl})
or as the action density of the tree-level improved L\"uscher-Weisz
action. We have shown that the integration of this Zeuthen flow
equation and the evaluation of classically improved observables do not
produce any $\mathcal O(a^2)$ effects to any order in the coupling
or, indeed, non-perturbatively. At this point it is important to remark 
that although the analysis has been performed in the context of pure
gauge theories, due to the classical nature of the flow equation,
the aforementioned results are still valid in QCD or if any number of
fermions in any representations are coupled to our gauge field. In the
particular case of the pure gauge theory the only $\mathcal O(a^2)$
effects originate either from the 4-dimensional 
lattice action or from the additional counterterm parameterized by
$c_b$ in the modified initial condition~(\ref{eq:cb}). Tree-level
$\mathcal O(a^2)$ improvement is achieved with the L\"uscher-Weisz
gauge action and $c_b=0$. 

We have explicitly checked that the proposed Zeuthen
flow equation does not generate any $\mathcal O(a^2)$ contribution to
tree level for a variety of gluonic observables (different
observables in arbitrary volumes). In doing so, we have shown that 
other proposals of the literature to improve the gradient flow 
(i.e.~the $\tau$-shift in~\cite{Cheng:2014jba}, the coefficients
in~\cite{Fodor:2014cpa} or the chair flow in~\cite{Ramos:2014kka}) 
in fact do produce $\mathcal O(a^2)$ effects in some of the considered
observables. In this sense, these proposals only produce vanishing 
$\mathcal O(a^2)$ cutoff effects in some particular situations
(i.e. $\langle E(t,x)\rangle$ in infinite volume), and this
cancellation should be regarded as accidental, and not as
improvement. 

Our results can be extended in various directions. 
First, it appears straightforward to extend the classical
$a$-expansion to further observables, for example the 
the energy-momentum tensor. When considering $n$-point correlation
functions of such observables with $n>1$ or if boundary conditions do
not respect translation invariance in some directions (as is the case
with SF and open boundary conditions), some additional work is
required to also eliminate total derivative terms which may contribute at  
any order in $a$. We also note that the improvement of observables and
the flow equation are conceptually separate from the $\mathcal O(a^2)$
effects at $t=0$. It is therefore conceivable 
to push the expansion further, in order to also cancel terms at
$\mathcal O(a^4)$. It is not clear how complicated this 
would be for the flow equation, but it is certainly an option for
observables. However, one should 
be aware that higher order improvement would typically render these
observables less local in lattice units. 
Another natural generalization would be the extension of our work to
include fermions and the fermionic flow equation, 
introduced in ref.~\cite{Luscher:2013cpa}.


\section*{Acknowledgments}
The authors want to show special gratitude to R.~Sommer for his help
and advice in many steps of this work. In the course of this work we
have benefited from discussions with 
M.~Garc\'{i}a Perez, A.~Gonz\'alez-Arroyo, M.~L\"uscher,  A.~Patella,
S.~Schaefer and our colleagues in the
ALPHA-collaboration. S.S.~is grateful for the hospitality 
extended to him at DESY-Zeuthen where this project has been initiated, and 
to both the CERN theory group and the Yukawa Institute for Theoretical
Physics (programme YITP-T-14-03), where essential progress was
made. The authors want to thank the organizers of the
workshop ``High-precision QCD at low energy'' and the staff of the
``Centro de ciencias Pedro Pascual'' in Benasque for the nice
atmosphere that we needed to finish this work. We warmly thank
A. Portelli for providing beers in the very last stage of this work.

S.S.~is  partially supported by Science Foundation Ireland under
grant 11/RFP/PHY3218. 


\appendix

\section{Conventions and notation}
\label{ap:conv}

We will use the summation convention for colour indices
\begin{equation}
 a,b,\ldots = 1,\dots, N^2-1 \,,
\end{equation}
 but not for space-time indices $\mu,\nu, \ldots$, as this may lead to
confusion in the discussion of lattice artefacts. Trace over color
indices will be denoted by ${\rm tr}$ (lower case), while trace over Lorentz
indices will be denoted with the symbol ${\rm Tr}$ (upper case).

$SU(N)$ gauge fields live in the Lie algebra $\mathfrak{su}(N) $ and
are traceless antihermitian $N\times N$ matrices. Any element $X\in
\mathfrak{su}(N)$ of this 
algebra can be written as $X = X^a T^a$ where the components $X^a$ are
real numbers and the generators $T^a$ are themselves antihermitian
$N\times N$ matrices chosen to obey the normalization  
\begin{equation}
  \tr(T^aT^b) = -\frac12\delta_{ab}\,.
\end{equation}

On the lattice the links $U_\mu(x)$ belong to the gauge group
$SU(N)$. For an arbitrary function of the link variables $f(U_\mu(x))$, the
Lie-algebra valued derivative is given by
\begin{displaymath}
  \partial_{x,\mu} f(U_\mu(x)) = T^a \partial_{x,\mu}^a f(U_\mu(x)) = 
  T^a\frac{{\rm d}}{{\rm d}\epsilon}f(e^{\epsilon T^a}U_\mu(x))
  \bigg|_{\epsilon=0}\,. 
\end{displaymath}

Fourier transformations on an infinite lattice with lattice spacing
$a$ are defined as 
\begin{eqnarray}
  A_\mu(x) &=& \int_p e^{\imath px+\imath p_\mu a/2}\tilde A_\mu(p)\,,
\end{eqnarray}
where 
\begin{equation}
  \int_p = \int_{-\pi/a}^{\pi/a} \frac{d^4p}{(2\pi)^4}\,.
\end{equation}
On a hypercubic lattice of volume $L^4$ we define 
\begin{equation}
  A_\mu(x) = \frac{1}{L^4} \sum_n e^{\imath px+\imath
    p_\mu a/2}\tilde A_\mu(p)\,,   
\end{equation}
with $p_\mu=2\pi n_\mu/L$ and $n_\mu = 0,\dots,L/a-1$. It is
convenient to introduce the lattice derivatives
\begin{eqnarray}
  \partial_\mu \phi(x) &=& \frac{\phi(x+a\hat \mu) - \phi(x)}{a}\,, \\
  \partial_\mu^* \phi(x) &=& \frac{\phi(x) - \phi(x-a\hat \mu)}{a}\,,
\end{eqnarray}
and also the covariant derivatives given by
\begin{eqnarray}
  a\nabla_\mu f(x) &=& U_\mu(x)f(x+a\hat\mu)U_\mu(x)^\dagger - f(x)\,, \\
  a\nabla_\mu^* f(x) &=& f(x) - U_\mu(x-a\hat \mu)^\dagger f(x-a\hat\mu)U_\mu(x-a\hat\mu)\,. \\
\end{eqnarray}

Along the work we use the following definitions of lattice momenta 
\begin{eqnarray}
  \hat p_\mu &=& \frac{2}{a}\sin(ap_\mu/2)\,, \\
  \mathring p_\mu &=& \frac{1}{a}\sin(ap_\mu)\,, \\
  \hat c_\mu &=& \frac{1}{a}\cos(ap_\mu/2) \,.
\end{eqnarray}


\section{Absence of odd powers of $a$ in the classical expansion $S_\text{fl}[V]$}
\label{ap:odd}

In this appendix we demonstrate that the apparent presence of
odd powers of $a$ in the classical expansion~(\ref{eq:FmuTseries})
is an artefact of the way the expansion was set up. In particular,
we will show that Symanzik's effective action for the flow action only contains
terms which are even powers of $a$.

\subsection{Re-exanding around the midpoint of the link}

Indeed, the expansion about $x$ does not account for the fact that
the equation is derived for a given link variable $V_\mu(t,x)$,
relating the lattice 
points $x$ and $x+a\hat\mu$. Odd powers of $a$ in the expansion are
due to this asymmetric 
treatment, as can be shown explicity to all orders in $a$ for the LHS
of the flow 
equation, Eq.~(\ref{eq:flow_LHS}).
First, we define the unitary matrices $\Omega_\mu(t,x)$ as the
parallel transporters along the half link 
from $x$ to the midpoint $\tilde{x} = x+ \frac12 a\hat\mu$,
\begin{equation}
  \Omega_\mu(t,x) = {\cal P} \exp\left\{ a \int_{\frac12}^{1} du\,
    B_\mu\left(t,z(u)\right)\right\} 
  \label{eq:halflink}
\end{equation}
i.e.~compared to the path ordered exponential $V_\mu(t,x)$
(\ref{eq:Vpathord}) we here only 
integrate over second half of the path parameterizing the link.
Now we can perform the parallel transport to the midpoint $\tilde{x}$,
defining 
\begin{eqnarray}
  F_\mu(t,x) &=& \Omega_\mu(t,x) \tilde{F}_\mu(t,\tilde{x})
                 \Omega_\mu(t,x)^{-1}, \\ 
\end{eqnarray}
and, analogously, $\tilde{L}(t,\tilde{x})$, such that the term in the
lattice flow action density, 
\begin{equation}
  {\cal L}^{(\mu)} = \tr\{ L_\mu(t,x) F_\mu(t,x) \} =  \tr \{
  \tilde{L}_\mu(t,\tilde{x})\tilde{F}_\mu(t,\tilde{x}) \}, 
\end{equation}
can be expressed in terms of fields defined at the midpoint. To obtain
the expansion in powers 
of $a$ about $\tilde{x}$  one may simply re-expand the expansion about
$x$ obtained previously. 
The parallel transporters $\Omega_\mu(t,x)$ then merely render the
Taylor expansion covariant. 
Proceeding in this way yields, for the first term of
$\tilde{F}_\mu(t,\tilde{x})$, 
\begin{equation}
  a^{-1}\Omega_\mu(t,x)^\dagger \left[\partial_t
    V_\mu(t,x)\right]V_\mu(t,x)^\dagger \Omega_\mu(t,x) 
= \sum_{n=0}^\infty \frac{1}{(2n+1)!}
\left(\frac{a}{2}D_\mu\right)^{2n}\partial_t B_\mu(t,\tilde{x}), 
\end{equation}
which explicitly contains even powers of $a$ only. For the gradient
force term in 
$F_\mu(t,x)$ we have only worked out the first few orders of the
$a$-expansion explicitly. 
Therefore, the re-expansion cannot be carried out to all orders in $a$
and it is thus advisable to resort to some more general argument based
on symmetries. 
 
\subsection{Reflection symmetries}

We now consider the flow action in Eq.~(\ref{eq:flact}), but restricted to
plaquette and rectangle terms, 
as this is sufficient to discuss the case of the Zeuthen flow.
We consider a coordinate reflection ${\mathcal R}_\alpha$ in direction $\alpha$.
The point with coordinates $x_\mu$ transforms into $x'_\mu$ with
\begin{subequations}
\begin{equation}
 {\mathcal R}_\alpha:\qquad x_\mu \longrightarrow x'_\mu =
    \begin{cases}
      -x_\alpha, & \text{if $\mu = \alpha$},\cr
          x_\mu, & \text{if $\mu \ne \alpha$},
    \end{cases}
\end{equation}
The gauge field transforms under $\mathcal R_\alpha$
\begin{equation}
  V_\mu(t,x) \longrightarrow
    \begin{cases}
       V_\alpha(t,x' - a\hat \alpha)^\dagger, &  \text{if $\mu = \alpha$},\cr
       V_\mu(t,x'), &  \text{if $\mu \ne \alpha$},
     \end{cases}
\end{equation}
One may then show that the gradient force terms (the RHS of the flow
equation), transform for plaquette and rectangle terms, as follows:
\begin{equation}
   X_\mu(t,x) \rightarrow \begin{cases}
      -V_\alpha(t,x'-a\hat{\alpha})^\dagger X_\alpha(t,x'-a\hat{\alpha})
       V_\alpha(t,x'-a\hat{\alpha}) & \text{if $\mu=\alpha$},\cr
       X_\mu(t,x') & \text{if $\mu\ne\alpha$}.
      \end{cases}
  \label{eq:RalphaX}
\end{equation}
\end{subequations}
In fact, the same transformation behaviour is found for the left hand
side of the 
flow equation, so that Eq.~(\ref{eq:RalphaX}) equally holds for $F_\mu(t,x)$
of Eq.~(\ref{eq:Fmudef}). Hence, if the same transformation behaviour
(\ref{eq:RalphaX}) 
is imposed on the Lagrange multiplier field, $L_\mu(t,x)$,
we obtain for the different parts of the action density,
\begin{equation}
   \tr\left(L_\mu(t,x) F_\mu(t,x)\right) \rightarrow
     \begin{cases}
     \tr\left(L_\alpha(t,x-a\hat{\alpha}) F_\alpha(t,x'-a\hat{\alpha})\right)   & \text{if $\mu=\alpha$},\cr
       \tr\left(L_\mu(t,x') F_\mu(t,x')\right)& \text{if $\mu\ne\alpha$}.
      \end{cases}
  \label{eq:R0transfact}
\end{equation}
In particular, the action is invariant under such a reflection, as the only effect
consists in a re-ordering of the terms in the sum over the
$x_\alpha$-coordinate\footnote{One may think of the infinite lattice as a limiting case of
finite lattices with periodic boundary conditions where the sum over $x_\alpha$ is finite
and the re-ordering of terms in the sum is unproblematic.}.

\subsection{Example: reflection $\mathcal R_\alpha$ of the Wilson gradient force}

It is instructive to derive Eq.~(\ref{eq:RalphaX}) for 
the case of the plaquette action in some detail. The gradient force
in this case has the form,
\begin{equation}
  X_\mu(t,x) = \sum_\nu \left(P_{\mu\nu}(t,x) + Q_{\mu\nu}(t,x)^\dagger\right)_{\rm AH}
\end{equation}
and we need to distinguish the two cases $\mu=\alpha$ and $\mu\ne\alpha$.
Starting with $\mu=\alpha$ and  setting $y = x'- a \hat \alpha$
we obtain the transformation behaviour of these plaquettes
\begin{subequations}
\begin{equation}
  P_{\alpha\nu}(t,x) \longrightarrow  
   V_\alpha(t,y)^\dagger P_{\alpha\nu}(t,y)^\dagger V_\alpha(t,y),
\end{equation}
and, similarly,
\begin{equation}
  Q_{\alpha\nu}(t,x)^\dagger \longrightarrow
   V_\alpha(t,y)^\dagger Q_{\alpha\nu}(t,y) V_\alpha(t,y).
\end{equation}
\end{subequations}
Summing both expressions and taking the antihermitian part we thus obtain
\begin{equation}
   \left(P_{\alpha\nu}(t,x) + Q_{\alpha\nu}(t,x)^\dagger\right)_{\rm AH} \longrightarrow
   - V_\alpha(t,y)^\dagger\left(
    P_{\alpha\nu}(t,y) + Q_{\alpha\nu}(t,y)^\dagger \right)_{\rm AH} V_\alpha(t,y)\,,
\end{equation}
where we have used the relation $(M^\dagger)_{\rm AH} = -(M)_{\rm AH}$, valid
for any square matrix $M$.

Next we consider the case $\mu\ne\alpha$. The transformations of the plaquettes in this case
read
\begin{subequations}
  \begin{equation}
    P_{\mu\nu} (t,x) \longrightarrow
       \begin{cases}
        Q_{\mu\alpha}(t,x')^\dagger, & \text{if $\nu = \alpha$}, \cr
        P_{\mu\nu}(t,x'),      & \text{if $\nu \ne \alpha$},
      \end{cases}
  \end{equation}
  and
   \begin{equation}
    Q_{\mu\nu} (t,x)^\dagger\longrightarrow
       \begin{cases}
        P_{\mu\alpha}(t,x'), & \text{if $\nu = \alpha$}, \cr
        Q_{\mu\nu}(t,x')^\dagger,      & \text{if $\nu \ne \alpha$}.
      \end{cases}
  \end{equation}
\end{subequations}
Hence, Eq.~(\ref{eq:RalphaX}) follows and the part of the lattice flow action containing
the Wilson gradient force is indeed invariant under a reflection ${\mathcal R}_\alpha$.
We have also verified that this remains true for any gradient force obtained
from lattice actions containing both plaquettes and rectangles,
such as the L\"uscher-Weisz action.

\subsection{Lattice vs.~continuum reflections}

We now consider a total reflection, $\mathcal R = \mathcal R_0 \mathcal R_1 \mathcal R_2 \mathcal R_3$
of all space-time coordinates, i.e.
\begin{equation}
   \mathcal R: \quad x \longrightarrow x' = -x
\end{equation}
The part of the flow action density for fixed index $\mu$ then
transforms as follows:
\begin{equation}
   \tr\left\{L_\mu(t,x) F_\mu(t,x)\right\} \longrightarrow
   \tr\left\{L_\mu(t,x'-a\hat\mu) F_\mu(t,x'-a\hat\mu)\right\}.
\end{equation}
It is not difficult to see that the O($a$) offset in this transformation is
again an artefact of the asymmetric treatment of the links.
In fact, defining again the midpoint $\tilde{x} = x+ \frac12 a\hat\mu$, and
using the transformation of the "half link variables", $\Omega(t,x)$~(\ref{eq:halflink}),
\begin{equation}
 \mathcal R: \qquad \Omega_\mu(t,x) \longrightarrow \Omega_\mu\left(t,x'-\tfrac{a}2\hat\mu\right)^\dagger,
\end{equation}
we find that the transformation behaviour of the fields at the midpoint
is given by
\begin{eqnarray}
 \tilde{L}_\mu(t,\tilde{x}) &\longrightarrow& -\tilde{L}_\mu(t,-\tilde{x}), \\
 \tilde{F}_\mu(t,\tilde{x}) &\longrightarrow&  -\tilde{F}_\mu(t,-\tilde{x}),
\end{eqnarray}
i.e.~the reflection $\mathcal R$, once expressed in terms of the fields at the midpoint $\tilde{x}$
takes the same form as its continuum counterpart.
Therefore, the $a$-expansion of the corresponding part of the lattice flow action cannot
generate terms that are odd under $\mathcal R$, i.e. any term
\begin{equation}
  T_{\mu_1,\mu_2,\ldots,\mu_n}(t,\tilde{x}) \longrightarrow (-1)^n T_{\mu_1,\mu_2,\ldots,\mu_n}(t,-\tilde{x}),
\end{equation}
with an odd number $n$ of Lorentz indices can be excluded.
This together with the observation that all Lorentz vectors ($D_\mu$, $L_\mu$, $\partial_t L_\mu$,\ldots)
have odd canonical dimension, implies that any term containing an even number of them must
be even dimensional and thus be accompanied by an even power of $a$.

Treating the parts of the lattice flow action density with other values of $\mu$
in the same way, no odd powers of $a$ can be generated in the expansions
about the respective midpoints of the links relating $x+a\hat\mu$ and $x$.
As these midpoints coalesce to a single point in the continuum limit
this establishes this property for the $a$-expansion of the complete lattice flow action, for
gradient force terms containing plaquette and rectangle terms.

For these considerations to extend to the Zeuthen flow we only need to
check that the correction term,
\begin{equation}
   \nabla_\mu^\ast\nabla_\mu^{} X_\mu(t,x),
\end{equation}
transforms like $X_\mu(t,x)$ itself under the reflection $\mathcal R$. This is indeed the case,
so that the absence of odd powers of $a$ is confirmed for the Zeuthen flow, too.
Finally, while it is plausible that these considerations extend to
gradient force terms derived from lattice gauge actions
containing the ``chairs" and ``parallelograms", we did not check this
explicitly, as it is not needed for the discussion of the O($a^2$)
improved Zeuthen flow.

\section{Action and heat kernels to O($a^2$)}
\label{ap:gauge}

\begin{table}
  \centering
  \begin{tabular}{l|l}
  \textbf{Discretization} & $K_{\mu\nu}(p;\lambda)$ \\
\hline
  Plaquette & $\hat p^2\delta_{\mu\nu} -
(1-\lambda)\hat p_\mu\hat p_\nu$ \\
  L\"uscher-Weisz & $\hat p^2\delta_{\mu\nu} -
(1-\lambda)\hat p_\mu\hat p_\nu 
   + \frac{a^2}{12}\left[ (\hat p^4 + \hat p^2\hat p_\mu^2)\delta_{\mu\nu} - 
   \hat p_\mu \hat p_\nu (\hat p_\mu^2 + \hat p_\nu^2)\right]$\\
Clover & $\mathring p^2\hat c_\mu^2\delta_{\mu\nu} - 
  \mathring p_\mu\hat c_\mu \mathring p_\nu\hat c_\nu$\\
  Zeuthen & $(1-a^2\hat p_\mu^2/12)\left\{\hat p^2\delta_{\mu\nu} -
\hat p_\mu\hat p_\nu 
   + \frac{a^2}{12}\left[ (\hat p^4 + \hat p^2\hat p_\mu^2)\delta_{\mu\nu} - 
   \hat p_\mu \hat p_\nu (\hat p_\mu^2 + \hat
p_\nu^2)\right]\right\}+ \lambda\hat p_\mu\hat p_\nu$\\
 \hline
 \hline
  \textbf{Discretization} & $R_{\mu\nu}(p;\lambda)$ \\
 \hline
  Plaquette & $ -
  \frac{1}{12}\left[p^4\delta_{\mu\nu} - (1-\lambda)\frac{1}{2}p_\mu
  p_\nu(p_\mu^2 + p_\nu^2)\right]$ \\
  L\"uscher-Weisz & $\frac{1}{12}p^2p_\mu^2\delta_{\mu\nu} 
  - \frac{1+\lambda}{24}p_\mu p_\nu(p_\mu^2+p_\nu^2)$\\
Clover & $-\left(\frac{1}{3}p^4 +
    \frac{1}{4}p^2p_\mu^2\right)\delta_{\mu\nu} +
  \frac{7}{24}p_\mu p_\nu(p_\mu^2+p_\nu^2)$\\
  Zeuthen & $\frac{1}{24}p_\mu p_\nu\left[
(1+\lambda)p_\mu^2 - (1-\lambda)p_\nu^2
 \right]$\\
 \hline
  \end{tabular}
  \caption{Kernels $K_{\mu\nu}$ corresponding to different choices of
    discretization, and discretization effect corrections $R_{\mu\nu}$
    for some of the most popular choices. See
    appendix~\ref{ap:conv} for any unexplained notation.}
  \label{tab:kernel}
\end{table}

\subsection{Free lattice actions and their kernels}

The choice of observable, action and flow at tree level can be
parameterized by the kernels of free lattice actions, i.e.~the gauge action expanded to
second order in the gluon fields, possibly supplemented by a gauge fixing term. If a generic
lattice action with Wilson loops of length 4 and 6 is chosen then these are parameterized by a set of
coefficients $c_i$, $i=0,1,2,3$. An alternative is provided by directly inserting
the clover leaf defintion of the gluon field strength tensor gives
into a continuum  like action density. In momentum space any of these actions is written
\begin{equation}
  S_g[U;\{c_i\}] = \frac12 \int_p A_\mu^a(-p)
  K_{\mu\nu}(p;\lambda; \{c_i\}) A_\mu^a(p) + \mathcal O(g_0)
\end{equation}
Note that for the case of a
finite volume with twisted boundary conditions, the expressions of the
kernels $ K_{\mu\nu}(p;\lambda)$ are unchanged, but the integrals over
momenta have to be substituted by sums and the momentum has to be
interpreted as the sum of the space and color momentums (see
Eq.~(\ref{eq:ptw}) and the subsequent discussion). Gauge 
fixing is performed in any kernel by adding the usual gauge fixing 
term
\begin{equation}
K_{\mu\nu}(p;\lambda) = K_{\mu\nu}(p;0) + \lambda \hat{p}_\mu \hat{p}_\nu\,.
\end{equation}

Expanding the kernels to $\mathcal O(a^2)$ around their common
continuum limit, 
\begin{equation}
 K(p;\lambda) = K^{\rm cont}(p;\lambda) + a^2 R(p,\lambda)+\rmO(a^4)
 \label{eq:Kexpansion}
\end{equation}
with
\begin{equation}
 K^{\rm cont}_{\mu\nu}(p;\lambda) = p^2 \delta_{\mu\nu} +(\lambda-1) p_\mu p_\nu
\end{equation}
the leading cutoff effects are encoded in the structure of
$R_{\mu\nu}(p;\lambda)$. For example for a generic action made of an
arbitrary linear combination of loops of 4 and 6 links Eq.~(\ref{eq:latac})
we have
\begin{eqnarray}
\nonumber
  K^{(G)}(p) &=& \hat p^2\delta_{\mu\nu} - (1-\lambda)\hat p_\mu \hat p_\nu
  - a^2(c_1-c_2-c_3)\left[ 
    (\hat p^4 + \hat p^2\hat p_\mu^2)\delta_{\mu\nu} - \hat p_\mu\hat
    p_\nu(\hat p_\mu^2+\hat p_\nu^2)
  \right] \\
  &-& a^2(c_2+c_3)\left[ 
    (\hat p^2)^2\delta_{\mu\nu} - \hat p^2\hat p_\mu\hat p_\nu
  \right] \,,
\end{eqnarray}
and 
\begin{eqnarray}
  \nonumber
  R^{(G)}_{\mu\nu} &=& -\left[ \left(\frac{1}{12}+c_1-c_2-c_3\right)p^4
    + (c_1-c_2-c_3)p^2p_\mu^2 + (c_2+c_3)(p^2)^2\right]\delta_{\mu\nu} \\
  &+& p_\mu p_\nu \left[\left(\frac{1-\lambda}{24}
      +c_1-c_2-c_3\right)(p_\mu^2+p_\nu^2) + (c_2+c_3)p^2 \right]\,.
\end{eqnarray}

The expressions for $K_{\mu\nu}$ and $R_{\mu\nu}$ for the other common
choices of discretizations are written in table~\ref{tab:kernel}. Note
that the Zeuthen flow equation, even if it is not derived from the
gradient of an action, can also be parametrized to tree-level by a
kernel. The main difference is that the property
\begin{equation}
K_{\mu\nu}(p) = K_{\nu\mu}(p)\,,
\end{equation}
that is obeyed by any kernel derived from the gradient of an action
(i.e.~a consequence of the action being real) does not hold for the
Zeuthen flow.

\subsection{Heat kernels and propagators to $\mathcal O(a^2)$}

Given a kernel $K$ with arbitrary value of the gauge parameter $\lambda$ (or $\alpha$ in case of the flow kernel),
we would like to obtain the $a^2$ correction term to either the propagator,
i.e.~the inverse of $K$,
\begin{equation}
   D(p;\lambda) = K(p;\lambda)^{-1},
\end{equation}
or the heat kernel
\begin{equation}
  H(t,p;\alpha) = \exp\left(-t K(p;\alpha)\right),
\end{equation}
given the expansion of the kernel Eq.~\eqref{eq:Kexpansion}.

Starting with the propagator, we can formally invert,
\begin{equation}
  D = \left[K^{\rm cont} + a^2 R+\mathcal O(a^4) \right]^{-1} =
      \left[ \unit + a^2 D^{\rm cont} R +\mathcal O(a^4)\right]^{-1} D^{\rm cont},
\end{equation}
and then expand in $a^2$ to obtain
\begin{equation}
  D(p;\lambda) = D^{\rm cont}(p;\lambda)
  -a^2 D^{\rm cont}(p;\lambda) R(p;\lambda) D^{\rm cont}(p;\lambda) + \mathcal O(a^4).
\end{equation}

To work out the $\mathcal O(a^2)$ piece of the heat kernel we define
the transverse and longitudinal projectors
\begin{equation}
  T_{\mu\nu}(p) = \delta_{\mu\nu} - \frac{p_\mu p_\nu}{p^2},\qquad
  L_{\mu\nu}(p) =  \frac{p_\mu p_\nu}{p^2},
\end{equation}
in terms of which 
\begin{equation}
   K^{\rm cont}_{\mu\nu}(p;\alpha) = p^2 \left[T_{\mu\nu}(p) + \alpha L_{\mu\nu}(p) \right]\,,
\end{equation}
and the heat kernel in the continuum is given by
\begin{equation}
  H^{\rm cont}(t,p;\alpha) =  \exp\left(-t K^{\rm cont}(p;\alpha)\right)
  =  e^{-tp^2}T(p) + e^{-\alpha tp^2}
  L(p). 
\end{equation}
Note that $K^{\rm cont}(p;\alpha)$ and $R(p;\alpha)$ do
in general not commute. Nevertheless, it is not difficult to work out
the expansion  to $\mathcal O(a^2)$ (i.e. to first order in
$R(p;\alpha)$). Inserting 
(\ref{eq:Kexpansion}) in the exponent one obtains 
\begin{equation}
   e^{-tK(p;\alpha)} = e^{-tp^2} e^{u L(p) + v R(p;\alpha)} +
   \rmO(v^2), \qquad u=(1-\alpha)p^2 t,\qquad v=-a^2 t \,.
\end{equation}
Then, noting, that for $n>1$,
\begin{equation}
  (uL +vR)^n = u^n L + u^{n-1} v \left\{LR+(n-2)LRL +RL\right\} + \rmO(v^2)\,,
\end{equation}
the exponential series can be resummed with the result
\begin{equation}
  e^{u L + v R} = \sum_{n=0}^\infty \frac{(uL+vR)^n}{n!}
  = T + e^u L+v\left[ R-\overline{R} + \frac{e^u-1}{u}\left(\bar{R} +
      u LRL\right)\right] + \rmO(v^2)\,, 
\end{equation}
where
\begin{equation}
  \bar{R} = LR + RL -2LRL.
\end{equation}
The result for the heat kernel then is:
\begin{eqnarray}
  H(t,p;\alpha) &=& H^{\rm cont}(t,p,\alpha) + a^2 t
                    e^{-tp^2}\biggl\{\bar{R} -R  \nonumber\\ 
                  &&\, +\,\frac{1-e^{(1-\alpha)tp^2}}{(1-\alpha)tp^2}
                  \left[\bar R + (1-\alpha)tp^2 LRL\right]\biggr\} +
                     \rmO(a^4), 
\end{eqnarray}
where we have left out the arguments for the sake of readibility.
Note that the choice of Feynman gauge $\alpha=1$ for the heat kernel
is not a problem, as the apparent singularity 
cancels, with the result
\begin{equation}
   H(t,p,;1)= e^{-tp^2}\left\{1-a^2 t R(p;1) \right\} + \rmO(a^4).
   \label{eq:H_feynmangauge}
\end{equation}


\bibliography{/home/alberto/docs/bib/math,/home/alberto/docs/bib/campos,/home/alberto/docs/bib/fisica,/home/alberto/docs/bib/computing}

\end{document}